\documentstyle[12pt,epsf]{article}
\begin{document}

\begin{titlepage}
\title{A method of implementing Hartree-Fock calculations
with zero- and finite-range interactions}

\author{H. Nakada\footnote{E-mail: nakada@c.chiba-u.ac.jp}\\
{\it Department of Physics, Faculty of Science, Chiba University,}
\vspace*{-3mm}\\
{\it Yayoi-cho 1-33, Inage, Chiba 263-8522, Japan}\\
and\\
M. Sato\\
{\it Graduate School of Science and Technology, Chiba University,}
\vspace*{-3mm}\\
{\it Yayoi-cho 1-33, Inage, Chiba 263-8522, Japan}}

\date{\today}
\maketitle
\thispagestyle{empty}

\begin{abstract}
We develop a new method of implementing the Hartree-Fock calculations.
A class of Gaussian bases is assumed,
which includes the Kamimura-Gauss basis-set
as well as the set equivalent to the harmonic-oscillator basis-set.
By using the Fourier transformation
to calculate the interaction matrix elements,
we can treat various interactions in a unified manner,
including finite-range ones.
The present method is numerically applied
to the spherically-symmetric Hartree-Fock calculations
for the oxygen isotopes with the Skyrme and the Gogny interactions,
by adopting the harmonic-oscillator, the Kamimura-Gauss
and a hybrid basis-sets.
The characters of the basis-sets are discussed.
Adaptable to slowly decreasing density distribution,
the Kamimura-Gauss set is suitable to describe unstable nuclei.
A hybrid basis-set of the harmonic-oscillator
and the Kamimura-Gauss ones is useful to accelerate the convergence,
both for stable and unstable nuclei.
\end{abstract}


\vspace*{3mm}\noindent
\end{titlepage}

\pagestyle{plain}

\section{Introduction}
\label{sec:intro}

Since the invention of the secondary beam technology,
numerous experimental data on the unstable nuclei
have disclosed new aspects of the atomic nuclei.
Remarkable examples are the presence of the nuclear halos and skins,
and the dependence of magic numbers
on the neutron (or proton) excess~\cite{ref:magic}.
It should be noticed that both are closely related to the properties
of the single-particle (s.p.) orbits in the unstable nuclei.
In order to understand these new phenomena,
which have raised questions on some of our conventional picture
of the nuclear structure,
it is worthwhile reinvestigating the properties
of the s.p. orbits in nuclei.

Because the atomic nuclei are bound without an external field,
a mean-field is necessary to obtain the s.p. orbits.
The Hartree-Fock (HF) theory and its extensions,
which are self-consistent approaches,
will be a desirable tool
to investigate the s.p. orbits from microscopic standpoints.
In studying the structure of unstable nuclei,
it is a practical problem how to treat numerically the wave-functions
at relatively large $r$, because there may be a halo.
Most of the methods employed so far
are adapted to the nuclei with sharply decreasing densities
at the surface.
They are not necessarily eligible to reproduce the halo structure.
Furthermore, an important physics problem
lies in the effective interaction.
Not many effective interactions have been used
in the HF calculations of nuclei.
The Skyrme interaction~\cite{ref:VB72} has been popular
in the HF studies,
since the zero-range form is easy to be handled.
A number of parameter-sets have been proposed
for the Skyrme interaction.
In the Skyrme interaction the non-locality in the nuclear interaction
is approximated by the momentum dependence of the zero-range force.
This approximation was justified by Negele and Vautherin
via the density-matrix expansion~\cite{ref:NV72}.
However, despite the success for the stable nuclei
and its recent development~\cite{ref:HL98},
it has not been inspected sufficiently
whether the first few terms of the density-matrix expansion
give good description of the nuclei far from the $\beta$-stability.
In this regard, it is desired to deal also with finite-range interactions.
The Gogny interaction~\cite{ref:Gogny}
is the only finite-range interaction
widely applied to the mean-field calculations.
In almost all recent studies based on the Gogny interaction,
the D1S parameter-set~\cite{ref:D1S} is employed.
However, the D1S set has a problem
which is revealed in the unstable nuclei~\cite{ref:SNprep}.
It could be important to consider various possibilities
of the effective interactions in the mean-field calculations.

In this article, we develop a new method
to implement the HF calculations.
The following two points will be kept in mind:
(i) for the drip-line nuclei
the wave-functions in the asymptotic region
could be significant and therefore should be treated properly,
and
(ii) the method should have capability
of handling various effective interactions,
particularly some finite-range ones.
Satisfying these two conditions, the method developed in this paper
will be useful to study structure of unstable nuclei
within the HF framework.

\section{Single-particle bases}
\label{sec:basis}

In the following discussions we assume that
the s.p. orbits maintain the spherical symmetry,
for the sake of simplicity.
The extension of the method to the symmetry-breaking cases
will be straightforward.

The HF calculations are implemented
by solving the s.p. Schr\"{o}dinger equation ({\it i.e.} the HF equation)
iteratively.
There are two well-known ways
to solve the s.p. Schr\"{o}dinger equation.
One is to discretize the radial or spatial coordinate
with a finite mesh,
and to integrate the differential equation numerically.
The other is to introduce a basis-set
and to reduce the equation to an eigenvalue problem,
by applying the matrix representation to the s.p. Hamiltonian.
Unless the effective interaction has zero range,
the HF equation becomes an integro-differential equation
because the Fock term is non-local.
This makes the mesh method to be cumbersome.
On the contrary, we can store the two-body matrix elements
in the basis method, and then the non-locality in the Fock term
addresses no essential difficulty.
Since we would deal with finite-range interactions
as well as zero-range interactions,
it will be advantageous to introduce a certain set of s.p. bases.

Because dimensionality in practical calculations
is necessarily finite,
the calculated wave-functions more or less inherit
characters of the original bases.
Therefore the choice of the basis-set is important,
and could depend on the system under discussion, in general.
The harmonic-oscillator (HO) basis-set has been popular
in describing the s.p. orbits of nuclei.
However, while the HO set is indeed efficient in the stable nuclei,
this is not the case for the drip-line nuclei,
as will be shown in Section~\ref{sec:test}.
The density of the drip-line nuclei slowly decreases
for large $r$ (radial coordinate).
On account of the short-range character of the nuclear force,
the asymptotic form of the density distribution
should be exponential, $e^{-2\eta r}$~\cite{ref:NO94},
where $\eta=\sqrt{2ME}/\hbar c$ with the separation energy $E$.
In the drip-line nuclei, the density in the asymptotic region
could sizably contribute to physical quantities
such as the rms radius and the binding energy.
This is a sharp contrast to the stable nuclei. 
However, the exponential asymptotics is hardly expressed by the HO bases.
It is noticed that the exponent depends on $E$,
which is not obtained until the convergence in the HF calculation.
We have to reproduce not only the exponential form
but also the $E$ dependence of the exponent,
for the proper description of the asymptotics.
It is commented that, in the mesh method,
we need a large number of mesh points
to reproduce the density distribution in the asymptotic region,
as far as we keep the mesh size uniform.

We consider the s.p. bases having the following form,
\begin{eqnarray} \varphi_{\alpha\ell jm}({\mathbf r})
&=& R_{\alpha\ell j}(r)[Y^{(\ell)}(\hat{\mathbf r})\chi_\sigma]^{(j)}_m\,;
\nonumber\\
R_{\alpha\ell j}(r) &=& {\cal N}_{\alpha\ell j}
r^{\ell+2p_\alpha}\exp[-(r/\nu_\alpha)^2]\,.
\label{eq:basis} \end{eqnarray}
Here $Y^{(\ell)}(\hat{\mathbf r})$ expresses the spherical harmonics
and $\chi_\sigma$ the spin wave-function.
We drop the isospin index without confusion.
The index $\alpha$ indicates the extra power of $r$ ($p_\alpha$),
which is a non-negative integer,
and the range of the Gaussian ($\nu_\alpha$), simultaneously.
The constant ${\cal N}_{\alpha\ell j}$ is determined as
\begin{equation} {\cal N}_{\alpha\ell j}
= {2^{\ell+2p_\alpha+{7\over 4}}\over{\pi^{1\over 4}
\sqrt{(2\ell+4p_\alpha+1)!!}}}
\left({1\over\nu_\alpha}\right)^{\ell+2p_\alpha+{3\over 2}}\,,
\end{equation}
so as for $\langle\varphi_{\alpha\ell jm}|\varphi_{\alpha\ell jm}
\rangle$ to be unity.
Since the bases of Eq.~(\ref{eq:basis}) are non-orthogonal
between different $\alpha$'s,
the Schr\"{o}dinger equation leads to a generalized eigenvalue problem
when these bases are applied.
If we take $p_\alpha=0,1,2,\cdots$ with a constant range
$\nu_\alpha=\nu_\omega= \sqrt{2\hbar/M\omega}$,
the space spanned by these bases is equivalent to
that comprised of the HO bases.
Indeed, these bases coincide with the HO ones,
whose radial part is given by the associated Laguerre polynomials
of $2(r/\nu_\omega)^2$,
if the Gram-Schmidt orthogonalization is carried out
from the smaller $p_\alpha$ to the larger.
Because all the HO bases have the common Gaussian factor
$e^{-(r/\nu_\omega)^2}$,
the superposition of a limited number of the HO bases
has the asymptotic form of $e^{-(r/\nu_\omega)^2}$ again.
This is the reason why the HO basis-set fails
to describe the density (namely, the wave-function) asymptotics
of the drip-line nuclei.
On the other hand, Kamimura proposed a basis-set~\cite{ref:Kam88}
in which $\nu_\alpha$ is given by a geometric progression,
while keeping $p_\alpha=0$.
This set of bases,
which will be called Kamimura-Gauss (KG) basis-set in this article,
has been shown to work efficiently in few-body systems~\cite{ref:KKF},
including loosely bound ones.
Although each KG basis has the Gaussian asymptotics,
the exponential decrease of the density at large $r$
is appropriately described to a good approximation
by the superposition of the Gaussians with various ranges.
The present form of the bases (\ref{eq:basis}) covers
both the HO and KG bases.

The form of Eq.~(\ref{eq:basis}) allows wider variety of basis-sets
than the HO and the KG sets.
An immediate possibility is a hybridization of the HO and KG bases.
Another possibility may be stochastic selection
of $p_\alpha$ and $\nu_\alpha$,
though it is not explored in this paper.

The transformed harmonic-oscillator (THO) bases
were developed to reproduce the exponential asymptotics
in the density of the loosely bound nuclei,
and were applied to the mean-field calculations
with the Skyrme interaction~\cite{ref:THO}.
In order to obtain the energy dependence of the exponent,
the bases themselves are changed iteratively.
It was shown that the THO basis-set gives an improved description
of the properties of nuclei near the neutron drip-line,
over the HO set.
However, in Ref.~\cite{ref:THO} the actual calculation
seems to depend on the characteristics of the zero-range interaction.
It might not be easy to deal with finite-range interactions
by the THO bases.

The bases of Eq.~(\ref{eq:basis}) give the norm matrix of,
for each $(\ell,j)$,
\begin{eqnarray} N^{(\ell j)}_{\alpha\beta}
&=& \langle\varphi_{\alpha\ell jm}|\varphi_{\beta\ell jm}\rangle
\nonumber\\
&=& {{(2\ell+2p_\alpha+2p_\beta+1)!!}\over
\sqrt{(2\ell+4p_\alpha+1)!!(2\ell+4p_\beta+1)!!}}
\left({\nu_\beta\over\nu_\alpha}\right)^{p_\alpha-p_\beta}
\left({{2\nu_\alpha\nu_\beta}\over{\nu_\alpha^2+\nu_\beta^2}}
\right)^{\ell+p_\alpha+p_\beta+{3\over 2}}\,. 
\label{eq:norm}\end{eqnarray}
Under the $\ell$ and $j$ conservation,
the s.p. Hamiltonian matrix is given by
\begin{equation} h^{(\ell j)}_{\alpha\beta}
= \langle\varphi_{\alpha\ell jm}|\hat h
|\varphi_{\beta\ell jm}\rangle\,,
\label{eq:spham}\end{equation}
where $\hat h$ stands for the s.p. Hamiltonian.
Suppose that $|\psi_{n\ell jm}\rangle$ is a solution
of the s.p. Schr\"{o}dinger equation,
\begin{equation} \hat h |\psi_{n\ell jm}\rangle
= \epsilon_{n\ell j} |\psi_{n\ell jm}\rangle\,.
\label{eq:spSch}\end{equation}
By expanding $|\psi_{n\ell jm}\rangle$
by the bases $|\varphi_{\alpha\ell jm}\rangle$,
\begin{equation} |\psi_{n\ell jm}\rangle
= \sum_\alpha c^{(\ell j)}_{n,\alpha} |\varphi_{\alpha\ell jm}\rangle\,,
\label{eq:sp-expand}\end{equation}
the Schr\"{o}dinger equation (\ref{eq:spSch}) is represented
as the generalized eigenvalue problem,
\begin{equation}
\sum_\beta h^{(\ell j)}_{\alpha\beta} c^{(\ell j)}_{n,\beta}
= \epsilon_{n\ell j} \sum_\beta N^{(\ell j)}_{\alpha\beta}
 c^{(\ell j)}_{n,\beta}\,.
\end{equation}
Since the norm matrix $N^{(\ell j)}$ is real symmetric,
the Cholesky decomposition can be applied,
which is equivalent to the orthonormalization of the bases.
Then the generalized eigenvalue problem is converted
to the normal eigenvalue problem.

When we deal with non-orthogonal bases,
we have to be careful for the norm after the orthogonalization
not to be too small.
In particular,
the bases of Eq.~(\ref{eq:basis}) could compose an over-complete set,
as is obvious from the fact that the HO basis-set,
which is equivalent to the set of the bases having a fixed $\nu_\alpha$,
can already be complete.
If the norm after the orthogonalization were too small,
{\it i.e.} one of the bases were almost linearly-dependent
on the other bases,
a numerical instability could take place.
This condition may pose a practical limit on the present method
in choosing $p_\alpha$ and $\nu_\alpha$.

\section{Effective interaction}
\label{sec:effint}

The effective Hamiltonian for the nuclear mean-field theory
consists of the kinetic energy and the effective interaction,
\begin{equation}
H = K + V\,;\quad K = \sum_i {{\mathbf p}_i^2\over{2M}}\,,\quad
V = \sum_{i<j} v_{ij}\,.
\end{equation}
Here $i$ and $j$ are the indices of each nucleon.
The s.p. matrix element of the kinetic term is calculated as
\begin{eqnarray} \langle \varphi_{\alpha\ell jm}|
{{\mathbf p}^2\over{2M}}|\varphi_{\beta\ell jm}\rangle
\hspace*{10.5cm}\nonumber\\
= {1\over{2M \nu_\alpha \nu_\beta}} \left[
(2\ell+2p_\alpha+2p_\beta+3){{2\nu_\alpha \nu_\beta}\over
{\nu_\alpha^2+\nu_\beta^2}} - 2\left\{(\ell+2p_\alpha)
{\nu_\alpha\over\nu_\beta}+(\ell+2p_\beta)
{\nu_\beta\over\nu_\alpha}\right\}\right.
\nonumber\\
\quad \left. + 4{{(\ell+2p_\alpha)
(\ell+2p_\beta)+\ell(\ell+1)}\over{2\ell+2p_\alpha+2p_\beta+1}}
\cdot{{\nu_\alpha^2+\nu_\beta^2}\over{2\nu_\alpha \nu_\beta}}\right]
N^{(\ell j)}_{\alpha\beta}\,.
\label{eq:kin-me}\end{eqnarray}
It will be natural to assume the effective interaction $v_{ij}$
to be translationally invariant,
except for the density dependence mentioned below.

As stated in Section~\ref{sec:intro},
we would consider various types of the two-body interaction.
For the zero-range interaction like the Skyrme force,
the s.p. Hamiltonian is represented
in terms of the local densities and currents~\cite{ref:VB72,ref:CNP1}.
It is fast to compute the matrix elements of the s.p. Hamiltonian
via the local densities and currents.
However, this is not the case for finite-range interactions.
Hence we shall calculate the two-body interaction matrix elements
and store them,
as will be discussed in the subsequent section.

The saturation must be fulfilled in the nuclear HF approach.
This requires components
other than the momentum-independent two-body terms
in the central force~\cite{ref:FW}.
A density-dependent (or a three-body) interaction
is usually introduced.
Because the calculated density is renewed at each HF iteration,
it is impractical to store the matrix elements
of the density-dependent interaction.
We here assume the usual zero-range form
for the density-dependent interaction.
The contribution of this component to the s.p. Hamiltonian
is evaluated via the local densities.

In reproducing the saturation, it could be an alternative way
to introduce the momentum dependence in the central force,
which satisfies the translational invariance,
instead of the density dependence.
Although we do not consider that possibility in this paper
except for the case of the Skyrme interaction,
the momentum-dependent two-body interaction will be handled
in a similar manner to the momentum-independent interaction
discussed below.

In addition to the saturation properties
which are relevant to the central force,
the LS splitting is significant in the atomic nuclei.
This suggests necessity of the LS and/or the tensor forces,
though true origin of the LS splitting is still
under discussion~\cite{ref:LS}.
In most mean-field calculations so far,
the zero-range LS force was assumed~\cite{ref:VB72,ref:Gogny}.
The zero-range tensor force was sometimes taken into account
so as to cancel a certain term of the LS current~\cite{ref:S3}.
Finite-range LS and tensor forces are also considered,
as well as the zero-range ones.

We thus consider the effective interaction in the following form,
\begin{eqnarray} v_{12} &=& v_{12}^{\rm C} + v_{12}^{\rm LS}
 + v_{12}^{\rm TN} + v_{12}^{\rm DD}\,;\nonumber\\
v_{12}^{\rm C} &=& \sum_\mu (t_\mu^{\rm W} + t_\mu^{\rm B} P_\sigma
- t_\mu^{\rm H} P_\tau - t_\mu^{\rm M} P_\sigma P_\tau)
 f_\mu^{\rm C} (r_{12})\,,\nonumber\\
v_{12}^{\rm LS} &=& \sum_\mu (t_\mu^{\rm LSE} P_{\rm TE}
 + t_\mu^{\rm LSO} P_{\rm TO})
 f_\mu^{\rm LS} (r_{12})\,{\mathbf L}_{12}\cdot
({\mathbf s}_1+{\mathbf s}_2)\,,\nonumber\\
v_{12}^{\rm TN} &=& \sum_\mu (t_\mu^{\rm TNE} P_{\rm TE}
 + t_\mu^{\rm TNO} P_{\rm TO})
 f_\mu^{\rm TN} (r_{12})\, r_{12}^2 S_{12}\,,\nonumber\\
v_{12}^{\rm DD} &=& t_3 (1 + x_3 P_\sigma)
 [\rho({\mathbf r}_1)]^\alpha \delta({\mathbf r}_{12})\,,
\label{eq:effint}\end{eqnarray}
where $f_\mu$ represents an appropriate function,
$\mu$ stands for the parameter attached to the function
({\it e.g.} the range of the interaction),
and $t_\mu$ the coefficient.
As examples of $f_\mu(r_{12})$,
the delta, the Gauss and the Yukawa forms will be considered.
The relative coordinate is given
by ${\mathbf r}_{12}= {\mathbf r}_1 - {\mathbf r}_2$
and $r_{12}=|{\mathbf r}_{12}|$.
Correspondingly, we define
${\mathbf p}_{12}= ({\mathbf p}_1 - {\mathbf p}_2)/2$.
$P_\sigma$ ($P_\tau$) denotes the spin (isospin) exchange operator.
$P_{\rm TE}$ ($P_{\rm TO}$) is the projection operator
on the triplet-even (triplet-odd) two-particle state,
\begin{equation}
P_{\rm TE} = {{1+P_\sigma}\over 2}\,{{1-P_\tau}\over 2}\,,
\quad P_{\rm TO} = {{1+P_\sigma}\over 2}\,{{1+P_\tau}\over 2}\,.
\label{eq:proj_T}\end{equation}
Similarly, the projection operators on the singlet states
are defined by
\begin{equation}
P_{\rm SE} = {{1-P_\sigma}\over 2}\,{{1+P_\tau}\over 2}\,,
\quad P_{\rm SO} = {{1-P_\sigma}\over 2}\,{{1-P_\tau}\over 2}\,.
\label{eq:proj_S}\end{equation}
${\mathbf L}_{12}$ is the relative orbital angular momentum,
\begin{equation} {\mathbf L}_{12}= {\mathbf r}_{12}
\times {\mathbf p}_{12}\,,
\label{eq:L12}\end{equation}
${\mathbf s}_1$, ${\mathbf s}_2$ are the nucleon spin operators,
and $S_{12}$ is the tensor operator,
\begin{equation} S_{12}= 4\,[
3({\mathbf s}_1\cdot\hat{\mathbf r}_{12})
({\mathbf s}_2\cdot\hat{\mathbf r}_{12})
- {\mathbf s}_1\cdot{\mathbf s}_2 ]\,.
\label{eq:tensor}\end{equation}
The nucleon density is denoted by $\rho({\mathbf r})$.

The Skyrme and the Gogny interactions are included
in the present category of effective interactions.
In the case of the Skyrme interaction,
we use $f_\delta^{\rm C}(r_{12}) = \delta({\mathbf r}_{12})$
for the momentum-independent central force.
Discussions on the momentum-dependent and LS terms
will be given in Appendices~\ref{app:mom-dep} and \ref{app:LS}.
In the case of the Gogny interaction,
we set $f_\mu^{\rm C}(r_{12}) = \exp[-(\mu r_{12})^2]$.
The LS force has the same form as in the Skyrme interaction.
In both interactions, $v_{12}^{\rm TN}=0$ is assumed,
except for the counter-term to a part of the LS current.
In some of the recent parameterization
of the Skyrme interaction~\cite{ref:SkI}
the LS contribution is expressed
only in the density-functional form,
without explicit correspondence to the two-body interactions.
They are not expressed in the form of Eq.~(\ref{eq:effint})
and will not be considered in this paper.

\section{Calculation of two-body interaction matrix elements}
\label{sec:tbme}

We now discuss how to compute the matrix elements
of the two-body interactions,
$v_{12}^{\rm C}$, $v_{12}^{\rm LS}$ and $v_{12}^{\rm TN}$.
Their contribution to $h^{(\ell j)}_{\alpha\beta}$
in Eq.~(\ref{eq:spham}) is given by
\begin{equation}
\sum_{n'\ell'j'J} \langle \hat N_{n'\ell'j'}\rangle
{{2J+1}\over{(2j+1)(2j'+1)}} \sum_{\alpha'\beta'}
c^{(\ell'j')}_{n',\alpha'}\, c^{(\ell'j')}_{n',\beta'}\,
\langle(\alpha\ell j, \alpha'\ell'j')J|
(v_{12}^{\rm C} + v_{12}^{\rm LS} + v_{12}^{\rm TN})
|(\beta\ell j, \beta'\ell'j')J\rangle_{\rm A}\,
\label{eq:occapp}\end{equation}
where $\langle \hat N_{n\ell j}\rangle$ denotes
the occupation number of the s.p. orbit.
The expression $(\alpha\ell j)$ represents the basis
$|\varphi_{\alpha\ell j}\rangle$ of Eq.~(\ref{eq:basis})
in Section~\ref{sec:basis}.
Though not shown explicitly,
the proton-neutron degrees-of-freedom should be considered
in Eq.~(\ref{eq:occapp}), in practice.
As has been discussed in the preceding section,
the contribution of $v_{12}^{\rm DD}$ to the s.p. Hamiltonian
is calculated through the proton and neutron density distributions
at each iterative process, as in Ref.~\cite{ref:CNP1}.

Let us take the central force $v_{12}^{\rm C}$ as an example.
The anti-symmetrized two-body matrix element in Eq.~(\ref{eq:occapp})
is obtained from the non-anti-symmetrized ones,
\begin{equation}
\langle (j'_1 j'_2)J| v_{12}^{\rm C} |(j_1 j_2)J\rangle_{\rm A}
= \langle (j'_1 j'_2)J| v_{12}^{\rm C} |(j_1 j_2)J\rangle
- \langle (j'_1 j'_2)J| v_{12}^{\rm C} |(j_2 j_1)J\rangle\,,
\label{eq:antisym}\end{equation}
where $|~\rangle_{\rm A}$ ($|~\rangle$) denotes
anti-symmetrized (non-anti-symmetrized) state vector.
Without confusion, the symbol $j$ is regarded
as an abbreviation of $(\alpha\ell j)$
and the proton-neutron degrees-of-freedom.
As shown in Eq.~(\ref{eq:occapp}),
we only need the anti-symmetrized matrix elements
with $\ell_1=\ell'_1$, $\ell_2=\ell'_2$, $j_1=j'_1$ and $j_2=j'_2$
for the spherical HF calculations.
However, we here discuss how to evaluate the matrix elements
in more general manner,
as will be useful in the case that the symmetry is broken.
Inserting Eq.~(\ref{eq:effint}), we have
\begin{equation}
\langle (j'_1 j'_2)J| v_{12}^{\rm C} |(j_1 j_2)J\rangle
= \sum_\mu \langle (j'_1 j'_2)J| (t_\mu^{\rm W}
 + t_\mu^{\rm B} P_\sigma - t_\mu^{\rm H} P_\tau
 - t_\mu^{\rm M} P_\sigma P_\tau) f_\mu^{\rm C} (r_{12})
|(j_1 j_2)J\rangle\,.
\label{eq:tbcent}\end{equation}
Without loss of generality
the spatial function $f_\mu^{\rm C}(r_{12})$ is assumed
to be common for the Wigner, Bartlett, Heisenberg and Majorana terms.
The difference among them are in the spin and isospin operators,
for which we use the notation ${\cal O}_\sigma$
$(=1\mbox{ or }P_\sigma)$ and ${\cal O}_\tau$
$(=1\mbox{ or }P_\tau)$.
By converting the $jj$-coupling to the LS-coupling,
each term of the interaction in Eq.~(\ref{eq:tbcent}) is written as
\begin{eqnarray}
\langle (j'_1 j'_2)J| f_\mu^{\rm C} (r_{12}) {\cal O}_\sigma
{\cal O}_\tau |(j_1 j_2)J\rangle \hspace*{10.5cm}\nonumber\\
= \sum_{LS} (2L+1)(2S+1)\sqrt{(2j_1+1)(2j_2+1)(2j'_1+1)(2j'_2+1)}
\left\{\begin{array}{ccc}l_1&{1\over 2}&j_1\\ l_2&{1\over 2}&j_2\\
L&S&J\end{array}\right\}
\left\{\begin{array}{ccc}l'_1&{1\over 2}&j'_1\\ l'_2&{1\over 2}&j'_2\\
L&S&J\end{array}\right\}\nonumber\\
\quad \times \langle (l'_1 l'_2)L| f_\mu^{\rm C} (r_{12})
|(l_1 l_2)L\rangle\,\langle{\cal O}_\sigma\rangle_S\,
\langle{\cal O}_\tau\rangle\,, \hspace*{8cm}
\end{eqnarray}
where $\langle{\cal O}_\sigma\rangle_S$
($\langle{\cal O}_\tau\rangle$) denotes
the expectation value of ${\cal O}_\sigma$ (${\cal O}_\tau$)
of the spin (isospin) part.
The spatial part in the right-hand side is defined by
\begin{eqnarray}
\langle (l'_1 l'_2)L| f_\mu^{\rm C} (r_{12}) |(l_1 l_2)L\rangle
= \int d^3r_1 d^3r_2 R_{j'_1}(r_1) R_{j'_2}(r_2)
\{[Y^{(\ell'_1)}(\hat{\mathbf r}_1)
 Y^{(\ell'_2)}(\hat{\mathbf r}_2)]^{(L)}_M\}^*
\nonumber\\
\times f_\mu^{\rm C} (r_{12}) R_{j_1}(r_1) R_{j_2}(r_2)
[Y^{(\ell_1)}(\hat{\mathbf r}_1)
 Y^{(\ell_2)}(\hat{\mathbf r}_2)]^{(L)}_M\,.
\label{eq:tbinteg2}\end{eqnarray}

The spatial matrix element (\ref{eq:tbinteg2})
can straightforwardly be calculated for simple forms of the interaction
such as the delta form.
However, we here intend to handle various types of interactions,
including the Yukawa form.
For this purpose,
we utilize the Fourier transform of $f_\mu(r_{12})$,
as was exploited in Ref.~\cite{ref:HS61} for the HO bases.

The Fourier transformation of $f_\mu(r_{12})$ gives
\begin{equation} \tilde f_\mu (k)
= \int d^3r_{12}\,f_\mu(r_{12}) e^{-i{\mathbf k}\cdot{\mathbf r}_{12}}\,.
\label{eq:Four1}\end{equation}
By inverting this transformation, we obtain
\begin{equation} f_\mu(r_{12}) = {1\over{(2\pi)^3}}
\int d^3k \tilde f_\mu(k) e^{i{\mathbf k}\cdot{\mathbf r}_{12}}
= {1\over{(2\pi)^3}} \int d^3k \tilde f_\mu(k)
 e^{i{\mathbf k}\cdot{\mathbf r}_1} e^{-i{\mathbf k}\cdot{\mathbf r}_2}\,.
\label{eq:Four2}\end{equation}
Substituting Eq.~(\ref{eq:Four2}) into Eq.~(\ref{eq:tbinteg2}),
we find the ${\mathbf r}_1$ and ${\mathbf r}_2$ integrals
are separated at the expense of the ${\mathbf k}$ integration.
The angular integration is implemented by using
\begin{equation}
e^{i{\mathbf k}\cdot{\mathbf r}}
= 4\pi \sum_{\lambda} i^\lambda (2\lambda+1)
j_\lambda(kr)\,Y^{(\lambda)}(\hat{\mathbf k})
\cdot Y^{(\lambda)}(\hat{\mathbf r})\,,
\end{equation}
where $j_\lambda(x)$ denotes the spherical Bessel function,
deriving
\begin{eqnarray}
\langle (l'_1 l'_2)L| f_\mu^{\rm C} (r_{12}) |(l_1 l_2)L\rangle
= \sum_\lambda \sqrt{(2\ell_1+1)(2\ell_2+1)}\,
(\ell_1\,0\,\lambda\,0|\ell_1'\,0)
(\ell_2\,0\,\lambda\,0|\ell_2'\,0)\nonumber\\
\times \int_0^\infty k^2dk\,\tilde f_\mu^{\rm C}(k)\,
{\cal I}^{(0)}_1(k)\, {\cal I}^{(0)}_2(k)\,.
\end{eqnarray}
Here ${\cal I}^{(0)}_1$ and ${\cal I}^{(0)}_2$
are defined as
\begin{equation}
{\cal I}^{(0)}_i(k) = \int_0^\infty r^2dr
 j_\lambda(kr) R_{\alpha_i\ell_i j_i}(r) R_{\alpha'_i\ell'_i j'_i}(r)\,,
\label{eq:r-int-cent0}\end{equation}
with the subscript $i\,(=1,2)$ corresponds to the nucleon index,
which actually represents $(\alpha_i\ell_i j_i, \alpha'_i\ell'_i j'_i)$.
Since the radial part of the present basis $R_{\alpha\ell j}(r)$,
given in Eq.~(\ref{eq:basis}), has the Gaussian form,
${\cal I}^{(0)}(k)$ is calculated analytically,
\begin{equation}
{\cal I}^{(0)}(k) = \zeta_{\alpha\ell,\alpha'\ell'}\,
\left({k\over\kappa_{\alpha\alpha'}}\right)^\lambda\,
L^{(\lambda+{1\over 2})}_{{\ell+\ell'-\lambda\over 2}+p_\alpha+p_{\alpha'}}
\!\left[\left({k\over\kappa_{\alpha\alpha'}}\right)^2\right]
\,\exp\left[-\left({k\over\kappa_{\alpha\alpha'}}\right)^2\right] \,,
\label{eq:r-int-cent}\end{equation}
where $L^{(\alpha)}_n(x)$ is the associated Laguerre polynomial and
\begin{eqnarray}
\zeta_{\alpha\ell,\alpha'\ell'} &=&
{{\displaystyle 2^{{{\ell+\ell'}\over 2}+p_\alpha+p_{\alpha'}}\,
\Gamma\left({{\ell+\ell'-\lambda}\over 2}+p_\alpha+p_{\alpha'}+1\right)}
\over\sqrt{(2\ell+1)!! (2\ell'+1)!!}} \hspace*{2cm}\nonumber\\
&&\hspace*{0.5cm}\times
\left({\nu_{\alpha'}\over\nu_\alpha}\right)^{{\ell-\ell'\over 2}
+p_\alpha-p_{\alpha'}}
\left({{2\nu_\alpha \nu_{\alpha'}}\over{\nu_\alpha^2+\nu_{\alpha'}^2}}
\right)^{{\ell+\ell'+3\over 2}+p_\alpha+p_{\alpha'}}\,,\\
\kappa_{\alpha\alpha'} &=&
2 \sqrt{{1\over\nu_\alpha^2}+{1\over\nu_{\alpha'}^2}}\,.
\end{eqnarray}
We thus obtain the following expression
for the spatial part of the matrix element,
\begin{eqnarray}
\langle (l'_1 l'_2)L| f_\mu^{\rm C} (r_{12}) |(l_1 l_2)L\rangle
\hspace*{10cm}\nonumber\\
= \sum_\lambda \sqrt{(2\ell_1+1)(2\ell_2+1)}\,
(\ell_1\,0\,\lambda\,0|\ell_1'\,0)
(\ell_2\,0\,\lambda\,0|\ell_2'\,0)
\zeta_{\alpha_1\ell_1,\alpha'_1\ell'_1}
\zeta_{\alpha_2\ell_2,\alpha'_2\ell'_2}
\hspace*{2cm}\nonumber\\
\times \int_0^\infty k^2dk\,\tilde f_\mu^{\rm C}(k)\,
\left({k\over\kappa_{\alpha_1\alpha'_1}}\right)^\lambda
\left({k\over\kappa_{\alpha_2\alpha'_2}}\right)^\lambda
\, L^{(\lambda+{1\over 2})}_{{\ell_1+\ell'_1-\lambda\over 2}
+p_{\alpha_1}+p_{\alpha'_1}}
\!\left[\left({k\over\kappa_{\alpha_1\alpha'_1}}\right)^2\right]\,
\hspace*{1cm}\nonumber\\
\times
\, L^{(\lambda+{1\over 2})}_{{\ell_2+\ell'_2-\lambda\over 2}
+p_{\alpha_2}+p_{\alpha'_2}}
\!\left[\left({k\over\kappa_{\alpha_2\alpha'_2}}\right)^2\right]
\,\exp\left[-\left({k\over\kappa_{\alpha_1\alpha'_1}}\right)^2
-\left({k\over\kappa_{\alpha_2\alpha'_2}}\right)^2\right] \,.\quad
\label{eq:tbcrad}\end{eqnarray}

Whereas it is not easy in general to evaluate numerically
the multi-dimensional integrals to a high precision,
we have only one-dimensional $k$ integral
in Eq.~(\ref{eq:tbcrad}).
Moreover, even this $k$ integral is analytically carried out
for the typical interaction forms.
Recall that the associated Laguerre polynomial is defined as
\begin{equation}
L^{(\alpha)}_n(x) = \sum_{q=0}^n {{\Gamma(\alpha+n+1)}\over
{\Gamma(\alpha+q+1)\,(n-q)!}}\,{x^q\over{q!}}\,.
\end{equation}
The $k$ integral in Eq.~(\ref{eq:tbcrad})
turns out to be the sum of the integrals with the form
\begin{equation}
\int_0^\infty dk\,k^{2n+2} e^{-(k/\bar\kappa)^2} \tilde f_\mu(k)\,,
\label{eq:k-integ}\end{equation}
where $n$ is a certain integer
and $\bar\kappa^2=(1/\kappa_{\alpha_1\alpha'_1}^2
+1/\kappa_{\alpha_2\alpha'_2}^2)^{-1}$.
For the zero-range interaction
such as the momentum-independent term of the Skyrme interaction,
$f_\delta(r_{12})=\delta({\mathbf r}_{12})$ leads to
$\tilde f_\delta(k)=1$.
Here we substitute $\delta$ for the suffix $\mu$
to show the function form explicitly.
The integral of Eq.~(\ref{eq:k-integ}) therefore reduces to
\begin{equation}
\int_0^\infty dk\,k^{2n+2} e^{-(k/\bar\kappa)^2}
= {{(2n+1)!!\sqrt{\pi}}\over{2^{n+2}}}\,\bar\kappa^{2n+3}\,.
\end{equation}
For the Gaussian form such as the Gogny interaction,
$f_\mu(r_{12})=e^{-(\mu r_{12})^2}$ derives
$\tilde f_\mu(k)=({\sqrt\pi}/\mu)^3\, e^{-(k/2\mu)^2}$.
Thus the integral of Eq.~(\ref{eq:k-integ})
for the Gauss interaction is
\begin{equation} \left({\sqrt\pi\over\mu}\right)^3
\int_0^\infty dk\,k^{2n+2} e^{-(k/\bar\kappa)^2-(k/2\mu)^2}
= \left({\sqrt\pi\over\mu}\right)^3 {{(2n+1)!!\sqrt{\pi}}\over{2^{n+2}}}
\left({1\over{\bar\kappa^2}}+{1\over{4\mu^2}}\right)^{-(n+{3\over 2})}\,.
\end{equation}
In both cases the $k$ integral of (\ref{eq:k-integ})
yields an analytic function.
For the Yukawa interaction,
$f_\mu(r_{12})=e^{-\mu r_{12}}/\mu r_{12}$ leads to
$\tilde f_\mu(k)=4\pi/\mu(\mu^2+k^2)$.
The integration of (\ref{eq:k-integ}) is still written
in a compact form, by using the error function,
\begin{eqnarray} {4\pi\over\mu}
\int_0^\infty dk\,{{k^{2n+2}}\over{\mu^2+k^2}}\,e^{-(k/\bar\kappa)^2}
\hspace*{8cm}\nonumber\\
= {{2\pi^{3\over 2}\bar\kappa}\over\mu} (-\mu^2)^n
\left\{\sum_{r=0}^n (2r-1)!! \left(-{{\bar\kappa^2}\over{2\mu^2}}\right)^r
- {2\mu\over\kappa}\,e^{(\mu/\bar\kappa)^2}
\,{\rm Erfc}\!\left({\mu\over\bar\kappa}\right)\right\}\,,
\end{eqnarray}
where
\begin{equation} {\rm Erfc}(x) =
\int_x^\infty e^{-z^2} dz\,.
\end{equation}

As is shown in Appendix~\ref{app:mom-dep},
the momentum-dependent interaction,
such as contained in the Skyrme interaction,
can be handled in an analogous manner.
The treatment of the LS and the tensor forces are discussed
in Appendices~\ref{app:LS} and \ref{app:tensor}.
By the present technique we can deal with various interactions,
either zero-range or finite-range,
in a unified manner.

In coding a computer program,
we should prepare a subprogram for the integration
of Eq.~(\ref{eq:k-integ}).
This integral of Eq.~(\ref{eq:k-integ}) is the only part
dependent on the interaction form.
Therefore various interaction form can be handled
just by substituting the subprogram.
Moreover, it is unnecessary to carry out numerical integration
in calculating the interaction matrix elements,
for the delta, the Gauss and the Yukawa interactions.
Even for a more complicated form of the interaction,
numerical integration is only needed for Eq.~(\ref{eq:k-integ}),
as far as its Fourier transform $\tilde f_\mu(k)$ is known.

The above technique is also applicable to the Coulomb interaction.
The interaction form of $f(r_{12})=1/r_{12}$ yields
$\tilde f(k)=4\pi/k^2$,
and the $k$ integral of Eq.~(\ref{eq:k-integ}) becomes
\begin{equation} 4\pi \int_0^\infty dk\,k^{2n} e^{-(k/\bar\kappa)^2}
= (2n-1)!! \pi^{3\over 2}\,{{\bar\kappa^{2n+1}}\over{2^n}}\,.
\end{equation}
This is immediately obtained
from the $\mu\rightarrow 0$ limit in the Yukawa interaction.
Although the Coulomb exchange energy among the protons
was often approximated~\cite{ref:S3} as
\begin{equation}
E_{\rm exc}^{\rm Coul} \simeq -{3\over 4} e^2
\left({3\over\pi}\right)^{1\over 3}
\int [\rho_p({\mathbf r})]^{4\over 3}\,d^3r\,,
\end{equation}
this approximation can be lifted in the present approach.
It should be mentioned that an alternative method
for exact treatment of the Coulomb energy
was proposed recently~\cite{ref:AER01},
where the Coulomb force is transformed into an integration of Gaussians.

Once storing the interaction matrix elements
and having an estimate of $c^{(\ell j)}_{n,\alpha}$
in Eq.~(\ref{eq:sp-expand}),
we obtain the s.p. Hamiltonian from Eqs.~(\ref{eq:kin-me},\ref{eq:occapp}),
and via the densities for the contribution of $v_{12}^{\rm DD}$.
Solving the HF equation by iteration,
we can implement the HF calculation.

\section{Numerical tests}
\label{sec:test}

In this section we shall demonstrate the present method
via the HF calculations for the oxygen isotopes,
using both zero- and finite-range interactions.
The HF calculations are carried out
with maintaining the spherical symmetry and the parity conservation.
If the valence orbit is partially occupied,
the contribution of the orbit to the s.p. Hamiltonian
is averaged over the magnetic quantum numbers $m$,
as shown in Eq.~(\ref{eq:occapp}).
We first investigate the characters of the bases,
by taking the Skyrme interaction
with the SLy4 parameter-set~\cite{ref:SLy}.
Secondly the application to a finite-range interaction,
for which the Gogny D1S~\cite{ref:D1S} is used,
will be shown.
In both cases the Coulomb interaction is exactly treated,
as mentioned above.
The center-of-mass energies are corrected
approximately, by taking only the one-body kinetic part into account.

It takes longer time to implement HF calculations in heavy nuclei
than in light nuclei.
Though we restrict our application to the oxygen isotopes in this paper,
it is sufficiently practical to apply the present method
to the Pb isotopes, as will be shown in Ref.~\cite{ref:SNprep}.

\subsection{Selection of single-particle basis-sets}

We use several sorts of single-particle basis-sets,
composed of the bases having the form of Eq.~(\ref{eq:basis}).
Each of the bases is characterized by the index $\alpha$,
which actually corresponds to the parameters
$p_\alpha$ and $\nu_\alpha$.
In practical calculations,
we restrict the values of $p_\alpha$ and $\nu_\alpha$ to a certain extent;
otherwise there are too many possibilities.
As mentioned earlier, we can take both the HO-equivalent basis-set
and the KG basis-set, by choosing the parameters appropriately.
As well as these sets,
a hybrid basis-set will be tested,
in numerical calculations shown in the subsequent subsections.

The basis-set equivalent to the HO one is obtained
from Eq.~(\ref{eq:basis}), by posing
\begin{equation} p_\alpha=\alpha-1\,,\quad
\nu_\alpha = \nu_\omega\,,
\quad (\alpha=1,2,\cdots,K)
\label{eq:HO-par}\end{equation}
where
\begin{equation}
\nu_\omega = \sqrt{{2\hbar}\over{M\omega}}\,.
\label{eq:HO-range}\end{equation}
In the numerical calculations shown below,
we do not consider the nucleus dependence
of the $\nu_\omega$ parameter,
taking $\hbar\omega=41.2\times 24^{-1/3}\,{\rm MeV}$,
for the sake of simplicity.
After the Gram-Schmidt orthogonalization, the basis index $\alpha$
corresponds to the number of nodes $n$ in the HO bases.
In the usual calculations by the HO bases,
the truncation is made in terms of $N_{\rm sh}=2n+\ell$.
However, we here fix the number of the s.p. bases $K$
irrespective of $\ell$ and $j$,
to be fair with the case of the KG set mentioned below.
This indicates the truncation according to $n$,
rather than by $N_{\rm sh}$.
In the above set of (\ref{eq:HO-par}),
the maximum value of $n$ corresponds to $K-1$.

As stated in Section~\ref{sec:basis},
the bases of Eq.~(\ref{eq:basis}) are not orthogonal
and care must be taken so that the norm should not be too small
after the orthogonalization.
If we adopt the HO-equivalent set of Eq.~(\ref{eq:HO-par}),
the norm of the $K$-th basis appreciably decreases for growing $K$.
Numerical instability seems to occur for $K\geq 10$
in computations with the double precision.
Hence we always restrict ourselves to $K=7$
when we use the HO-equivalent basis-set.

The KG basis-set is obtained by
\begin{equation} p_\alpha=0\,,\quad
\nu_\alpha=\nu_1\,b^{\alpha-1}\,.\quad (\alpha=1,2,\cdots,K)
\label{eq:KG-par}\end{equation}
We use the same $\nu_1$, $b$ and $K$ for all $\ell$ and $j$.
If the common ratio $b$ is close to unity,
the overlap between the $\alpha$-th and the $(\alpha+1)$-th bases
is large.
Then the norm after the orthogonalization becomes vanishingly small,
which may lead to numerical instability.
On the other hand, if we adopt the larger value of $b$,
it is the more difficult to reproduce the wave-functions accurately.
For instance, in order to represent the smooth exponential decrease
of the density by a superposition of the Gaussians,
$b$ should not be very large.
In practice, the density distribution shows bumpy structure
for $b\geq 1.35$.
In the following calculations we fix $b=1.33$,
so as for the exponential asymptotics to be reproduced
in an effective manner.
For the range parameter $\nu_\alpha$,
we take one of them to be equal to $\nu_\omega$
in Eq.~(\ref{eq:HO-range}).
The $\nu_1$ value is determined accordingly;
for example, $\nu_1=\nu_\omega\,b^{-3}$ if we set $\nu_4=\nu_\omega$.

In the HF calculations we always confirm the convergence for iteration.
However, it is not easy in most cases to pursue the convergence
for increasing $K$.
This is also true for the KG basis-set.
The KG basis-set is characterized by three independent parameters;
the shortest range $\nu_1$, the longest range $\nu_K$
and the common ratio $b$.
Correspondingly, there are three courses
to increase the number of bases $K$.
One is to add the bases $\nu_{K+1}$ and so forth,
which have longer ranges than $\nu_K$,
with fixed $\nu_1$ and $b$.
The longer-range bases might be important to reproduce
the wave-functions in the asymptotic region,
particularly for the drip-line nuclei.
Another is to shorten $\nu_1$, keeping the longest range and $b$.
The shortest range $\nu_1$ is primarily relevant to the wave-functions
deeply inside the nucleus.
The other is to take smaller $b$.
This could be significant to accurate description
of the wave-functions in any region.
In order to attain the full convergence in the HF calculation,
all of the three courses should be tested.

While the KG set has an advantage in describing the wave-functions
in the asymptotic region,
it depends on the parameters how well the wave-functions
in the surface region is reproduced.
For the better description of the surface region by the KG set,
we usually need the smaller $b$.
An alternative way may be given by a hybridization of the KG set
and a small number of the HO-type bases.
In addition to the HO-equivalent and KG basis-sets,
we also test the hybrid basis-set.
Among the bases for each $\ell$ and $j$,
$(K-1)$ bases are taken to be the KG ones,
and for the last basis we use $p_\alpha=1$;
\begin{equation} \left\{\begin{array}{lll}p_\alpha=0\,,&
\nu_\alpha=\nu_1\,b^{\alpha-1}\,,& (\alpha=1,2,\cdots,K-1)\\
p_K=1\,,& \nu_K=\nu_\omega\,.&\end{array}\right.
\label{eq:hyb-par}\end{equation}
The parameters $b$ and $\nu_\omega$ are taken to be the same
as in the KG and HO bases mentioned above.

\subsection{Case of zero-range interaction}

We apply the present method to the even-$N$ oxygen isotopes.
Using the Skyrme SLy4 interaction,
we first compare results from the HO (the HO-equivalent, in practice),
the KG and the hybrid basis-sets for a fixed value of $K$; $K=7$.
For the KG and hybrid sets,
we take $\nu_1=\nu_\omega\,b^{-3}$,
leading to the longest range $\nu_K\cong 5.7\,{\rm fm}$
for the KG set
and $\nu_{K-1}\cong 4.3\,{\rm fm}$ for the hybrid set.

\begin{figure}
\epsfysize=9.0cm
\centerline{\epsffile{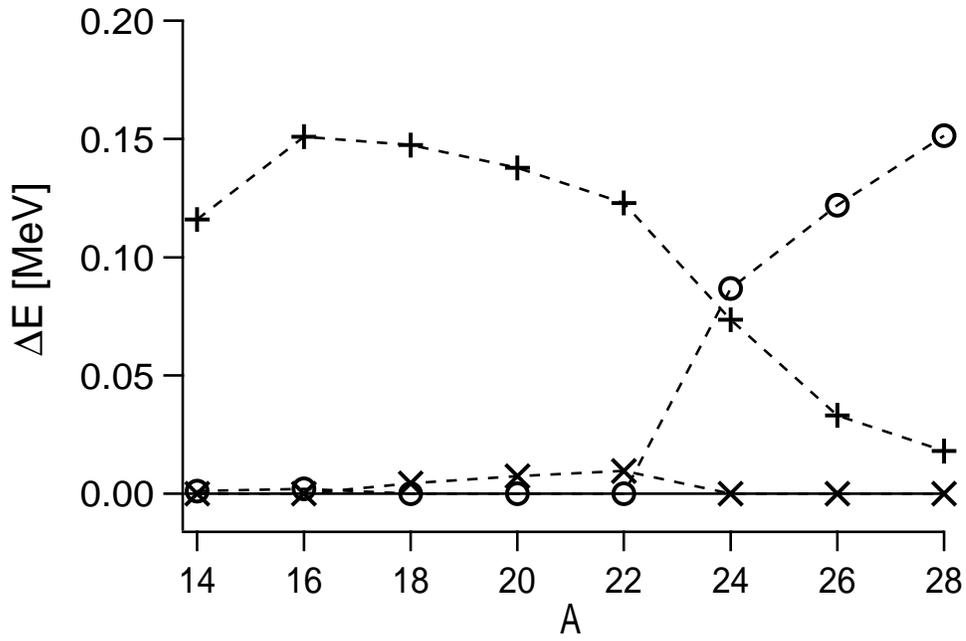}}
\vspace{2mm}
\caption{Comparison of the calculated HF energies
in the SLy4 interaction among the HO, the KG and the hybrid basis-sets
for the oxygen isotopes.
The circles, pluses and crosses indicate the energies
obtained from the HO, KG and hybrid sets, respectively.
In all cases $K=7$.
In each nucleus,
the energy differences are measured from the lowest HF energy
of those by the three basis-sets.
\label{fig:dE}}
\end{figure}

The variational character of the HF theory is available
in comparing results among different basis-sets.
As the total energy is lower,
it is closer to the true HF energy, in principle.
The total HF energies calculated with the HO, the KG
and the hybrid basis-sets are shown in Fig.~\ref{fig:dE},
where the energy differences from the lowest one
is plotted for each nucleus.
In comparison with the HO set,
the KG set gives higher energies for the $A\leq 22$ oxygen isotopes,
while in $A\geq 24$, where the neutron $1s_{1/2}$ orbit is occupied,
the KG set gives lower energies than the HO set.
This is ascribed to the broad radial distribution
of the $1s_{1/2}$ orbit,
which is hardly reproduced by the HO basis-set.
The hybrid basis-set works very well in the whole region.
The energies are close between the HO and the hybrid sets
in $^{14-22}$O, having the differences less than 0.01\,MeV,
and the hybrid set yields sizably lower energies
than the KG set for all of the calculated oxygen isotopes.
Thus the hybrid basis-set of Eq.~(\ref{eq:hyb-par})
is adaptable both to stable and unstable nuclei.

\begin{figure}
\epsfysize=17.0cm
\centerline{\epsffile{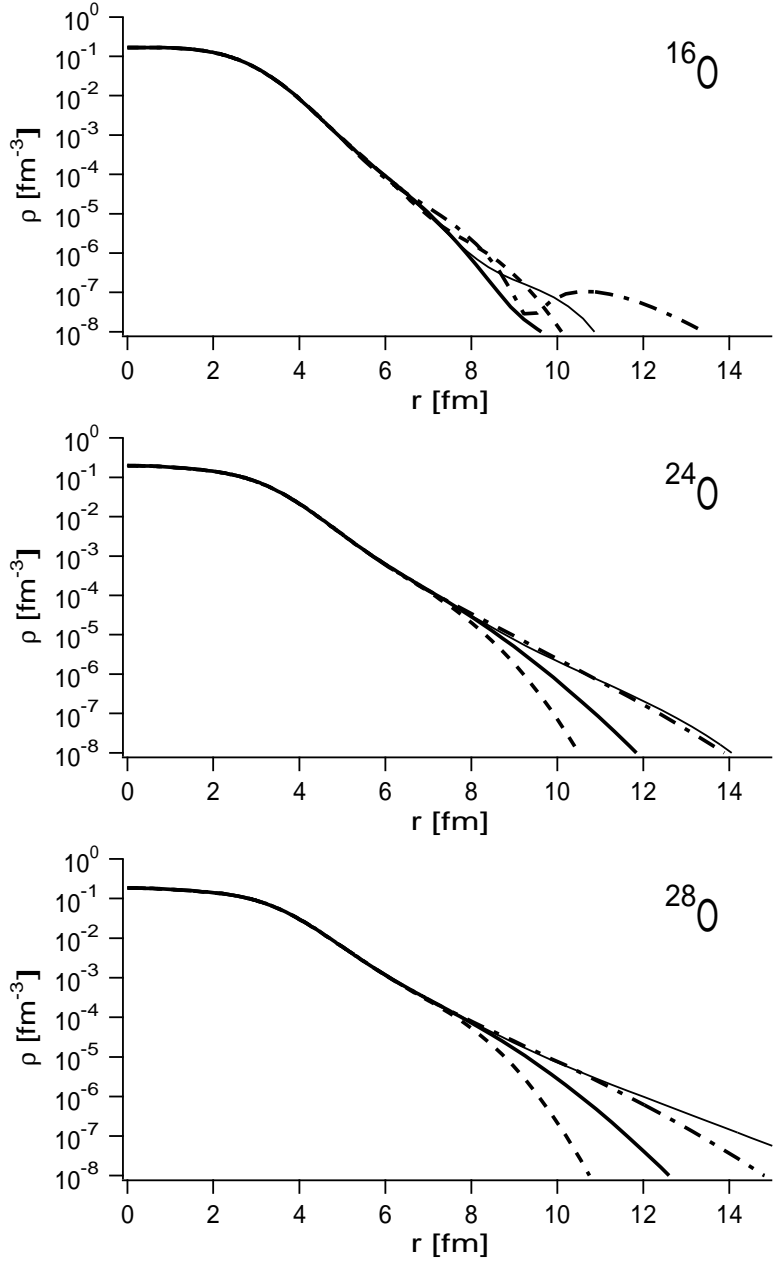}}
\vspace{2mm}
\caption{Comparison of the density distribution
in the SLy4 interaction among the HO, KG and hybrid basis-sets
for $^{16}$O, $^{24}$O and $^{28}$O.
The dashed, dot-dashed and thick solid lines are obtained
from the HO, KG and hybrid sets with $K=7$, respectively.
As a reference, the densities obtained with the $K=15$ hybrid set
are shown by the thin solid line.
\label{fig:rho_b7}}
\end{figure}

In Fig.~\ref{fig:rho_b7}, the density distribution is compared
among the three basis-sets, for $^{16}$O, $^{24}$O and $^{28}$O.
The density distribution in $r>6$\,fm tends to obey
to the exponential asymptotics.
Obviously, the density by the HO basis-set
does not distribute sufficiently as $A$ increases,
unable to reproduce the asymptotics for $^{24}$O and $^{28}$O.
The densities in $r>6\,{\rm fm}$ calculated with the HO set
behave quite analogously among the three nuclei,
suggesting that they originate in the character of the bases
and are not physical.
Therefore the HO set is practically incapable
of reproducing the asymptotics.
On the contrary, the KG and the hybrid basis-sets
reproduce the exponential asymptotics rather well.
Although the KG set does not give the exact exponential asymptotics,
it is possible to approximate the asymptotics
by the KG set in an effective sense.
The same holds for the hybrid set.
If the density becomes extremely low,
it is difficult to be reproduced by the KG set.
For this reason, the KG set shows fictitious behavior of the density
for $r>9$\,fm in $^{16}$O,
although the hybrid set yields rather smooth decrease.
For $^{24}$O and $^{28}$O,
the densities obtained by the KG set distribute more broadly
than those by the hybrid set.
This seems to caused by the difference in the longest range of the bases,
on which it depends how slowly decreasing density can be described.
Because we set the longest range to be $\nu_\omega\,b^3$
in the KG set while $\nu_\omega\,b^2$ in the hybrid one,
the KG set can reproduce broader distribution
that the hybrid set.

Figure~\ref{fig:rho_b7} shows the high adaptability of the KG basis-set,
particularly for the wave-functions in the asymptotic region.
The energy-dependent asymptotics are reproduced reasonably well
by a small number of bases.
Unlike the HO basis, an individual basis in the KG set
will not be a good first approximation of the nuclear s.p. wave-function.
Nevertheless, when a certain number of bases are superposed,
the KG set acquires remarkable flexibility in describing wave-functions.
In the nuclear wave-functions, the exponent of the asymptotic form
is energy dependent,
as is viewed in the nucleus dependence in Fig.~\ref{fig:rho_b7}.
The KG set (and therefore the hybrid set)
well approximates the exponentially damping wave-functions
with a moderate number of bases, whatever the separation energy is.

\begin{figure}
\epsfysize=17.0cm
\centerline{\epsffile{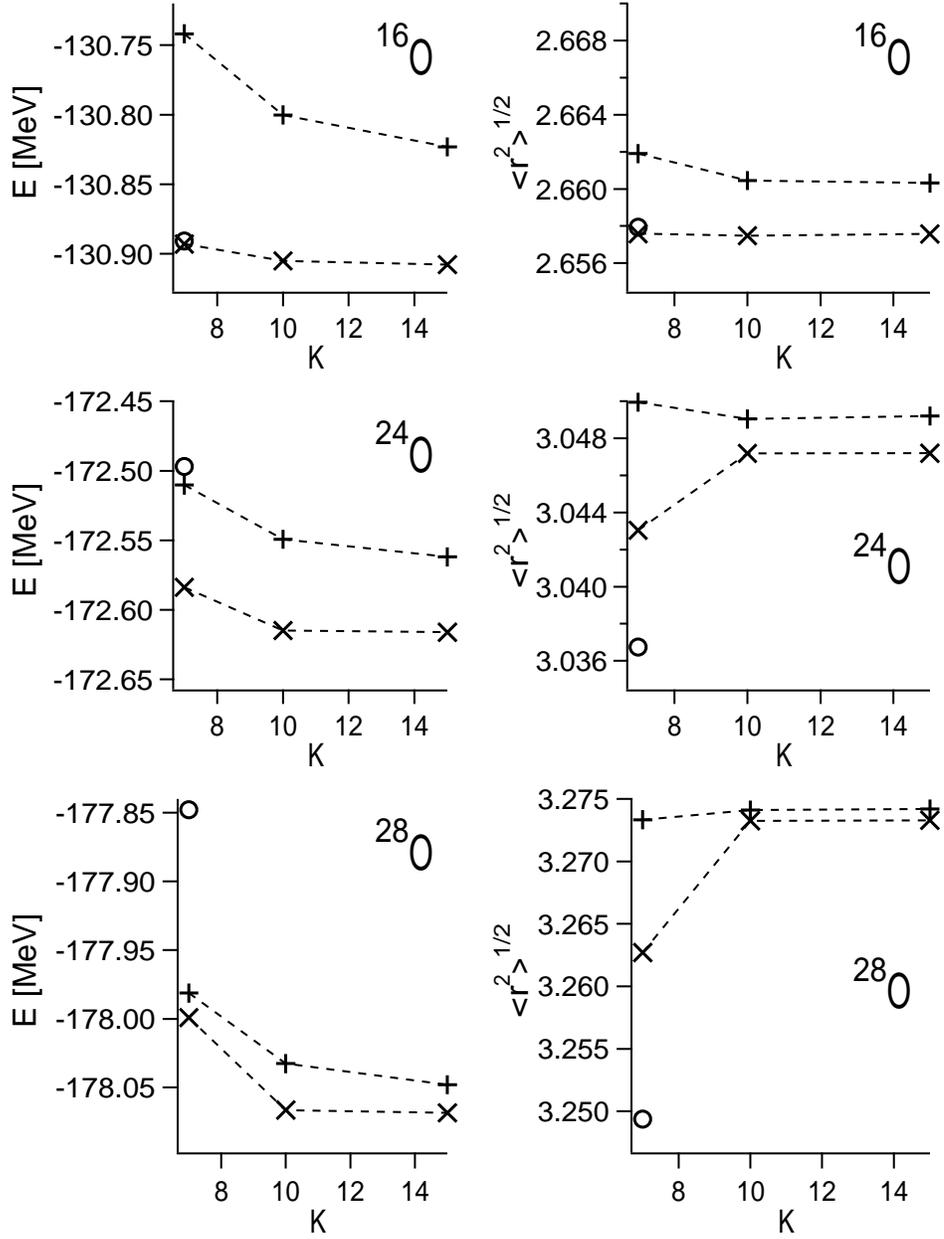}}
\vspace{2mm}
\caption{HF energies and rms matter radii
of $^{16}$O, $^{24}$O and $^{28}$O in the SLy4 interaction,
as a function of $K$ ({\it i.e.} number of the bases).
The circles, pluses and crosses represent
the results of the HO, KG and hybrid basis-sets, respectively.
The dashed lines are drawn to guide eyes.
\label{fig:conv}}
\end{figure}

We next increase $K$, the number of the bases for each $\ell$ and $j$.
The HF calculation is carried out using the KG and the hybrid basis-sets
for $K=10$ and $15$.
In the case of $K=10$, we take $\nu_1=\nu_\omega\,b^{-4}$
both for the KG and the hybrid sets,
giving the longest range $\nu_K\cong 10\,{\rm fm}$ for the KG
and $\nu_{K-1}\cong 7.5\,{\rm fm}$ for the hybrid set.
In the calculation with $K=15$, we take $\nu_1=\nu_\omega\,b^{-5}$
for the KG set and $\nu_\omega\,b^{-4}$ for the hybrid set,
both having the longest range of $31\,{\rm fm}$.
In Fig.~\ref{fig:conv},
the HF energies and the rms matter radii
are plotted as a function of $K$,
for $^{16}$O, $^{24}$O and $^{28}$O.
The density distributions obtained by the hybrid set of $K=15$
are already shown in Fig.~\ref{fig:rho_b7}.
For the matter radii, the center-of-mass correction is neglected,
corresponding to the density distributions
depicted in Fig.~\ref{fig:rho_b7},
in order to view the properties of the basis-sets directly.
Owing to the variational character,
the HF energies become lower as $K$ increases.
If we adopt the KG set,
the HF energies decrease slowly for increasing $K$.
On the contrary, the HF energies by the hybrid basis-set
are stable between $K=10$ and $15$,
where the biggest decrease among $^{14-28}$O
is merely 0.003\,MeV.
This sort of stability is also viewed in the rms matter radii.
Although the radii tend to be underestimated
by the hybrid set with $K=7$,
their difference between the $K=10$ and $15$ cases
are negligibly small; less than $1.5\times 10^{-4}\,{\rm fm}$
for all the $^{14-28}$O nuclei.
By comparing with the $K=15$ hybrid basis-set,
we view that the HO set provides reasonable values
of the HF energies and the matter radii
in the stable nucleus $^{16}$O,
while it is not satisfactory in the unstable ones
such as $^{24}$O and $^{28}$O.

Though one may think that the HF energies are convergent
in the $K=15$ result using the hybrid basis-set,
it does not imply the full convergence.
The HF energies further decrease, if we use smaller $b$.
Indeed, the HF energy becomes lower by about 0.01\,MeV
for a few of the oxygen isotopes, if we use $b=1.25$.
As in the calculations so far, it is quite a laborious work
to attain the full convergence in the HF calculation,
and is beyond the scope of this article.
In the same regard,
the HF energies shown in Figs.~\ref{fig:dE} and \ref{fig:rho_b7}
do not immediately mean a drawback of the KG set itself.
It is only indicated that the KG set with $b=1.33$
is not sufficient to describe the nuclear wave-functions
in the surface or the interior region.
If we use smaller $b$ than the present value,
the wave-functions around the surface may be reproduced
more accurately,
though it requires larger number of bases.
It will be fair to say that the convergence is accelerated
by using the hybrid set,
compared with the case of the KG set.

In the present approach,
the most time-consuming part in the numerical calculation
is the computation of the two-body interaction matrix elements.
It takes about 430\,sec of CPU time on HITAC SR8000
to calculate all the necessary matrix elements up to the $sd$-shell
when $K=7$,
although the program has not yet been tuned.
Moreover, the CPU time for computing the two-body matrix elements
is almost proportional to $K^4$.
On the other hand, the HF iteration need about 30\,sec for each nucleus,
and has dependence weaker than $K^2$.
Under this situation, a great advantage of the KG basis-set is that
it does not include parameters specific to mass number or nuclide.
Hence the same basis-set can be used for a number of nuclei.
With this advantage of the KG set,
we have to calculate the two-body interaction matrix elements only once
and do not have to recalculate them in systematic studies,
and thereby we can save the computation time.

\subsection{Case of finite-range interaction}

The present method of the HF calculation can be used
for finite-range interactions.
We demonstrate it via the calculation with the Gogny D1S interaction.

There is a problem in the Gogny D1S force
when it is applied to the mean-field calculations
using the KG basis-set~\cite{ref:SNprep}.
For the pure neutron matter with the D1S force,
the energy per nucleon diverges with the negative sign
at the high density limit.
Originating in this defect,
the HF energy goes to negative infinity in the finite nuclei,
when all the neutrons gather in the vicinity of the origin
({\it i.e.} the center-of-mass)
without overlap of the proton distribution.
For the $\beta$-stable nuclei,
this unphysical configuration is well separated
from the normal HF solution,
{\it i.e.} an energy minimum satisfying the saturation properties,
and the normal solution is stable enough to be obtained
in the numerical calculations.
However, it is not the case for the highly neutron-rich nuclei.
Even if the initial configuration is in the physical domain,
the tunneling to the unphysical configuration takes place
before convergence.
In order to circumvent the tunneling,
we need a certain cut-off of the high momentum components.
In the previous studies~\cite{ref:Gogny},
a sort of cut-off was implicitly introduced
by adopting a limited number of the HO bases.
On the other hand, we have to be cautious when we use the KG set.
The wave-function of $^{24}$O collapses via the tunneling
when bases having $\nu_\alpha\leq 1\,{\rm fm}$ are included.
Alternative to cutting off the high momentum components,
a way to avoid this problem is to modify the interaction parameters,
say, $x_3$ in $v_{12}^{\rm DD}$.
This possibility will be explored
in a forthcoming paper~\cite{ref:SNprep}.

In the numerical calculations,
we use the HO, the KG and the hybrid basis-sets
of (\ref{eq:HO-par}), (\ref{eq:KG-par}) and (\ref{eq:hyb-par}).
For the HO set, we take $K=7$ for each $\ell$ and $j$,
assuming the same value of $\nu_\omega$
as in the preceding subsection.
This corresponds to the $N_{\rm sh}\leq 13$ truncation
except for the $d$-orbits,
for which the $N_{\rm sh}=14$ bases are included,
and this basis-set is similar to that used
in most mean-field calculations with the Gogny interaction so far.
For the KG and hybrid sets,
we use $\nu_1=\nu_\omega\,b^{-2}$ and $b=1.33$,
{\it i.e.} $\nu_1=1.36\,{\rm fm}$, and $K=12$.

\begin{figure}
\epsfysize=17.0cm
\centerline{\epsffile{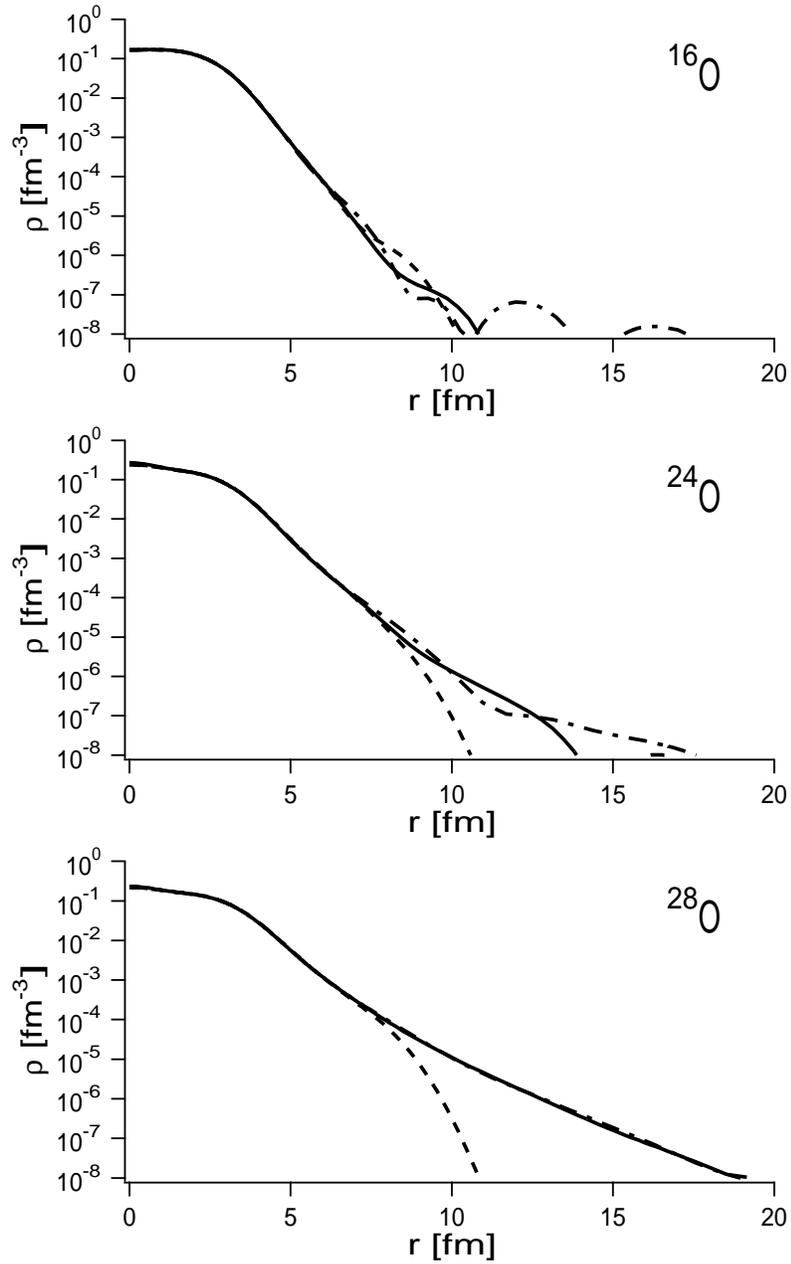}}
\vspace{2mm}
\caption{Comparison of the density distribution
in the D1S interaction
among the HO, KG and hybrid basis-sets for $^{16}$O,
$^{24}$O and $^{28}$O.
The dashed, dot-dashed and solid lines are obtained
from the HO, KG and hybrid sets, respectively.
\label{fig:rho_D1S}}
\end{figure}

The density distributions of $^{16}$O, $^{24}$O
and $^{28}$O are depicted in Fig.~\ref{fig:rho_D1S}.
The KG as well as the hybrid bases yield
the reasonable asymptotics for unstable nuclei,
whereas the HO set does not.
The variational character is not fully available
until establishing a rigorous cut-off scheme.
However, it is still informative to compare the HF energies,
as shown in Table~\ref{tab:eng_D1S}.
It is very likely that the energy of $^{28}$O by the HO set
is hampered by the ill asymptotic behavior,
being higher than the energy of the KG set.
As in the SLy4 case, the hybrid set gives the lowest energies
in all of $^{16}$O, $^{24}$O and $^{28}$O.
Thus the wave-functions in the asymptotic region,
which have a sizable contribution to the physical quantities
in highly neutron-rich nuclei,
might not be described properly
in the previous Gogny mean-field calculations by the HO bases,
even though the $\nu_\omega$ parameter is better tuned
than in the present calculation.

\begin{figure}
\epsfysize=9.0cm
\centerline{\epsffile{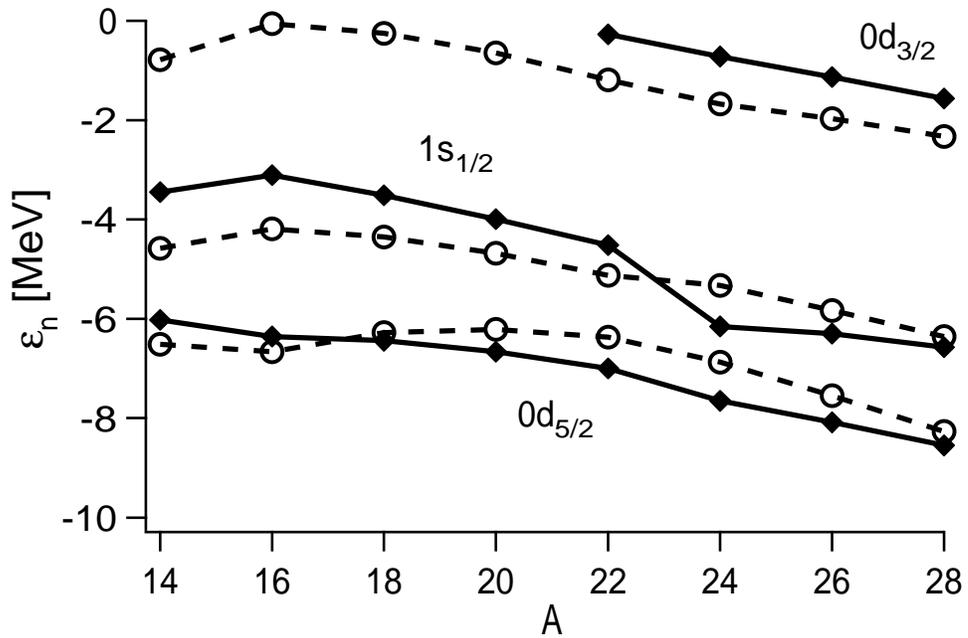}}
\vspace{2mm}
\caption{Neutron single-particle energies for the oxygen isotopes,
obtained from the HF calculations with the SLy4
(open circles) and with the D1S (diamonds) interactions.
The lines are drawn to guide eyes.
\label{fig:spe}}
\end{figure}

We next compare the HF results of the Gogny D1S interaction
with those of the Skyrme SLy4 interaction.
In Fig.~\ref{fig:spe}, the neutron s.p. energies
of the $sd$-shell orbits are shown for the oxygen isotopes.
For $^{14-20}$O, the $0d_{3/2}$ orbit is unbound in the D1S results,
and hence its energies are not presented.
The s.p. energies obtained from the SLy4 and the D1S interactions
are relatively close to each other.
A notable difference is found in the behavior of $\epsilon_n(1s_{1/2})$
around $^{24}$O;
the D1S interaction gives a kink at $^{24}$O, while the SLy4 does not.
When the $N$ (neutron number) dependence of the s.p. energies
in $^{16-22}$O is an effect of the occupation of $0d_{5/2}$,
changes in the s.p. energies from $^{22}$O to $^{24}$O
and from $^{24}$O to $^{28}$O are connected
to the $1s_{1/2}$ and $0d_{3/2}$ occupation.
It was argued based on the recent experimental data~\cite{ref:N16}
that $N=16$ becomes a magic number in the neutron-rich region.
The behavior of the s.p. energies around $^{24}$O could be
relevant to the magicity of $N=16$.
The kink in the D1S result gives rise to the relatively large gap
between $\epsilon_n(1s_{1/2})$ and $\epsilon_n(0d_{3/2})$
at $^{24}$O.
This would make the $^{24}$O core stiffer
than in the SLy4 interaction.
We have confirmed that the kink viewed in the Gogny D1S result
does not emerge in the results of other popular parameter-sets
of the Skyrme interaction, as well as in the SLy4 result.
Still it is not clear whether or not
the range of the interaction plays a role in this kink.
It is also commented that the kink at $^{24}$O
in the D1S interaction is not apparent
when we use the HO basis-set,
probably because of the wrong asymptotics.

\begin{figure}
\epsfysize=9.0cm
\centerline{\epsffile{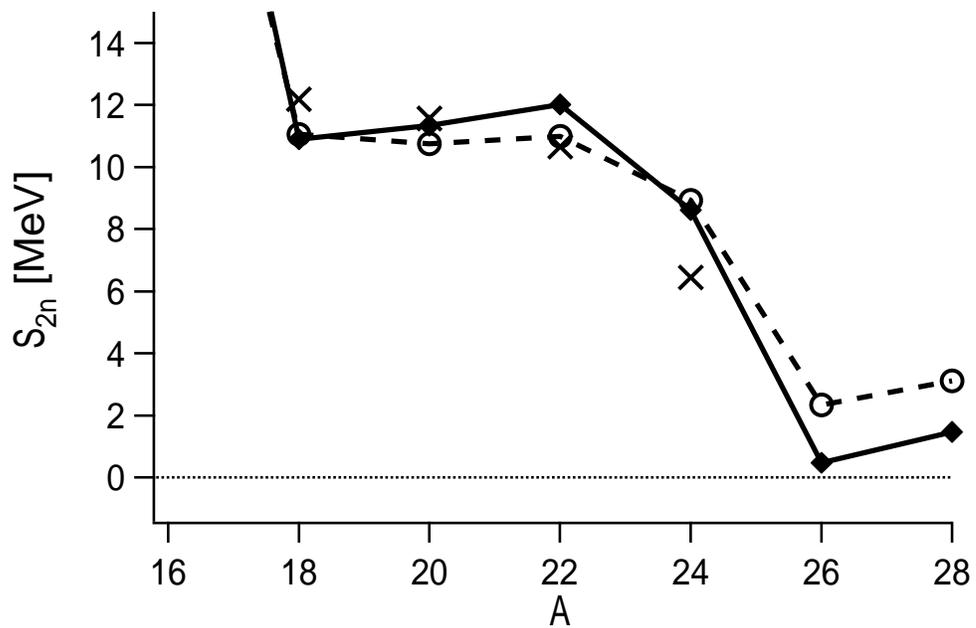}}
\vspace{2mm}
\caption{Two-neutron separation energies for the oxygen isotopes,
obtained from the HF calculations with the SLy4
(open circles) and with the D1S (diamonds) interactions.
The cross symbols represent the experimental data~\cite{ref:TI}.
\label{fig:S2n}}
\end{figure}

The two-neutron separation energy $S_{2n}$
is calculated as difference of the binding energies
between the neighboring isotopes.
The calculated values of $S_{2n}$ by the hybrid basis-set
($K=15$ for SLy4 and $K=12$ for D1S) are shown
and compared with the measured ones~\cite{ref:TI}
in Fig.~\ref{fig:S2n}.
The $S_{2n}$ values are not very different
between the SLy4 and D1S interactions for $^{18-24}$O.
It has been confirmed experimentally
that $^{26}$O and $^{28}$O are unbound~\cite{ref:O26,ref:O28}.
This indicates $S_{2n}<0$ for $^{26}$O.
This feature is not reproduced in the SLy4 result.
In the D1S interactions, we have slightly positive $S_{2n}$.
However, this seems to depend somewhat on the details
of the numerical set-ups;
for example, the treatment of the center-of-mass correction.
We just state that, in respect to $S_{2n}$ at $^{26}$O,
the D1S interaction gives preferable result to the SLy4 interaction.

In the present calculations
we do not take into account sufficiently the collectivity
due to the pairing correlations.
If we rely on the shell closure at $N=16$,
the pairing correlations will hardly change the energy of $^{24}$O,
while they lower the energies of $^{22}$O and $^{26}$O
to a certain extent.
Thus the pairing effects are expected to give
lower $S_{2n}$ at $^{24}$O and higher $S_{2n}$ at $^{26}$O
than the present HF values,
if we perform a Hartree-Fock-Bogolyubov calculation.

Concerning the computation time,
it is noted that the interaction dependence is weak
in the present method.
The CPU time for the Gogny interaction is almost the same
as in the Skyrme interaction.

\section{Summary and outlook}
\label{sec:summary}

We have developed a new method of the Hartree-Fock calculations.
This method has advantages
in reproducing the slowly decreasing density distributions
in unstable nuclei by a proper selection of bases,
and in treating various interactions including finite-range ones.

The key point of the method is adoption
of the Gaussian bases shown in Eq.~(\ref{eq:basis}).
This covers the Kamimura-Gauss (KG) basis-set,
as well as the basis-set equivalent
to the harmonic-oscillator (HO) one,
and may open wider variety.
We have also discussed a way to calculate
two-body interaction matrix elements,
by applying the Fourier transformation.
This is particularly suitable to the Gaussian bases
because the numerical integration can be avoided to a great extent.
Owing to this treatment of the interaction,
we can easily switch from an effective interaction to another.

The present method has numerically been tested
by using the Skyrme SLy4 and the Gogny D1S forces,
as representatives of the zero- and the finite-range interactions.
The calculations with the HO basis-set and with the KG basis-set
are compared.
It has been confirmed that the KG set efficiently
describes wave-functions in the asymptotic region
for the neutron-rich nucleus such as $^{24}$O and $^{28}$O.
When we adopt the KG set,
we do not have to change the bases from nucleus to nucleus,
since they do not contain nucleus-dependent parameters
like $\hbar\omega$.
Hence the KG basis-set is expected to be powerful
for systematic calculations.
We have also shown a way to improve the convergence
over the KG set.
If we use a hybrid basis-set, in which a HO-type basis
is added to the bases in the KG set,
the HF energies often decrease substantially.
The results on the s.p. energies and on $S_{2n}$
are also compared between the SLy4 and the D1S interactions,
for the oxygen isotopes.

The present method provides us with a useful tool to investigate
structure of the unstable nuclei,
particularly with finite-range interactions.
It will also be interesting to reconsider the effective interaction
within the mean-field approaches.
We can deal with various finite-range interaction,
not only for the central part, and even with the Yukawa form.
A research project in this line is under way.

While we have assumed the spherical symmetry in the discussions
in this paper,
it is straightforward to extend it to the deformed nuclei.
The future plan includes the extension
to the Hartree-Fock-Bogolyubov approach,
and the combination with the complex-scaling method
so as to handle the resonant single-particle orbits.
\\

\noindent
The authors are grateful to K. Kat\={o} and H. Kurasawa
for helpful discussions.
This work is supported in part
as Grant-in-Aid for Scientific Research (C), No.~13640263,
by the Ministry of Education, Culture, Sports, Science and Technology,
Japan.
Numerical calculations are performed on HITAC SR8000
at Information Processing Center, Chiba University.

\appendix
\section*{Appendices}

\section{Matrix elements of momentum-dependent part
of Skyrme interaction}
\label{app:mom-dep}

The central part of the Skyrme interaction is parameterized as
\begin{eqnarray} v_{12}^{\rm C}
&=& t_0(1+x_0 P_\sigma)\delta({\mathbf r}_{12})
+\frac{1}{2}t_1(1+x_1 P_\sigma)[\delta({\mathbf r}_{12}) {\mathbf p}_{12}^2
 +{\mathbf p}_{12}^2 \delta({\mathbf r}_{12})]\nonumber\\
&& +\,t_2(1+x_2 P_\sigma){\mathbf p}_{12} \cdot
     \delta({\mathbf r}_{12}) {\mathbf p}_{12} \,.
\end{eqnarray}
As has been discussed in Section~\ref{sec:tbme},
the two-body matrix elements of the $t_0$ term can be
treated in a unified way with the finite-range interactions.
We here show that, though the $t_1$ and $t_2$ terms
depend on the relative momentum ${\mathbf p}_{12}$,
they can also be treated in a similar manner
by using the Fourier transformation.

We first recall the identity
\begin{equation} \left[{\mathbf p}_{12}, \left[{\mathbf p}_{12},
\delta({\mathbf r}_{12})\right]\right]
= - \left(\nabla_{12}^2\,\delta({\mathbf r}_{12})\right)\,,
\end{equation}
where $\nabla_{12}=(\nabla_1-\nabla_2)/2$.
By using the Fourier transform of the delta function, we obtain
\begin{equation}
\left[{\mathbf p}_{12}, \left[{\mathbf p}_{12},
\delta({\mathbf r}_{12})\right]\right]
= [{\mathbf p}_{12}^2 \delta({\mathbf r}_{12})
+ \delta({\mathbf r}_{12}) {\mathbf p}_{12}^2]
- 2{\mathbf p}_{12}\cdot\delta({\mathbf r}_{12}){\mathbf p}_{12}
= {1\over{(2\pi)^3}} \int k^2 e^{i{\mathbf k}\cdot{\mathbf r}_{12}} d^3k
\,.
\end{equation}
Owing to the delta function,
$[{\mathbf p}_{12}^2 \delta({\mathbf r}_{12})
+ \delta({\mathbf r}_{12}) {\mathbf p}_{12}^2]$
vanishes when it operates on the spatially odd two-particle states.
Similarly,
$2{\mathbf p}_{12}\cdot\delta({\mathbf r}_{12}){\mathbf p}_{12}$
vanishes when acting on the spatially even states.
Hence we can write
\begin{eqnarray}
[{\mathbf p}_{12}^2 \delta({\mathbf r}_{12})
+ \delta({\mathbf r}_{12}) {\mathbf p}_{12}^2]
&=& {1\over{(2\pi)^3}} \int k^2 e^{i{\mathbf k}\cdot{\mathbf r}_{12}}
d^3k \cdot(P_{\rm SE}+P_{\rm TE})\,,\nonumber\\
-2{\mathbf p}_{12}\cdot\delta({\mathbf r}_{12}){\mathbf p}_{12}
&=& {1\over{(2\pi)^3}} \int k^2 e^{i{\mathbf k}\cdot{\mathbf r}_{12}}
d^3k \cdot(P_{\rm SO}+P_{\rm TO})\,.
\end{eqnarray}
The projection operators $P_{\rm SE}$, $P_{\rm TE}$, $P_{\rm SO}$
and $P_{\rm TO}$ can be incorporated in the spin-isospin part,
by using Eqs.~(\ref{eq:proj_T},\ref{eq:proj_S}).
Then the spatial matrix elements of
$[{\mathbf p}_{12}^2 \delta({\mathbf r}_{12})
+ \delta({\mathbf r}_{12}) {\mathbf p}_{12}^2]$
and $-2{\mathbf p}_{12}\cdot\delta({\mathbf r}_{12}){\mathbf p}_{12}$
are both evaluated by setting
$\tilde f_{\delta''}^{\rm C}(k)=k^2$
in Eq.~(\ref{eq:tbcrad}).

\section{Matrix elements of LS interaction}
\label{app:LS}

The LS interaction $v_{12}^{\rm LS}$ in Eq.~(\ref{eq:effint})
can be handled in a similar manner to the central force.
We here consider the non-anti-symmetrized matrix elements
of the LS force,
\begin{equation}
\langle (j'_1 j'_2)J| v_{12}^{\rm LS} |(j_1 j_2)J\rangle
= \sum_\mu \langle (j'_1 j'_2)J| (t_\mu^{\rm LSE} P_{\rm TE}
 + t_\mu^{\rm LSO} P_{\rm TO}) f_\mu^{\rm LS} (r_{12})
\,{\mathbf L}_{12}\cdot({\mathbf s}_1+{\mathbf s}_2)
|(j_1 j_2)J\rangle\,.
\label{eq:tbLS}\end{equation}
The LS force operates only on the spin-triplet two-particle states.
As in Appendix~\ref{app:mom-dep},
we separate $P_{\rm TE}$ and $P_{\rm TO}$,
denoting the projection operators
by ${\cal O}_{\sigma\tau} (=P_{\rm TE}\mbox{ or }P_{\rm TO})$
and their expectation values
by $\langle{\cal O}_{\sigma\tau}\rangle_{S=1}$.
It is noted that $f_\mu^{\rm LS}(r_{12})$
should not be the delta function,
since $\delta({\mathbf r}_{12}) {\mathbf L}_{12}=0$.

From the definition of Eq.~(\ref{eq:L12}),
${\mathbf L}_{12}$ is rewritten as
\begin{equation}
{\mathbf L}_{12} = {1\over 2}
(\mbox{\boldmath $\ell$}_1 + \mbox{\boldmath $\ell$}_2
- {\mathbf r}_1\times{\mathbf p}_2 - {\mathbf r}_2\times{\mathbf p}_1)\,,
\label{eq:L-sep}\end{equation}
where $\mbox{\boldmath $\ell$}_1 = {\mathbf r}_1\times{\mathbf p}_1$
and $\mbox{\boldmath $\ell$}_2 = {\mathbf r}_2\times{\mathbf p}_2$.
Since the $(\mbox{\boldmath $\ell$}_1+\mbox{\boldmath $\ell$}_2)$
operator does not change the spatial part of the wave-functions,
the matrix elements regarding
$(\mbox{\boldmath $\ell$}_1+\mbox{\boldmath $\ell$}_2)$
is handled in an analogous way to the central force,
\begin{eqnarray}
\langle (j'_1 j'_2)J| f_\mu^{\rm LS} (r_{12})
\,{1\over 2}(\mbox{\boldmath $\ell$}_1+\mbox{\boldmath $\ell$}_2)
\cdot({\mathbf s}_1+{\mathbf s}_2) {\cal O}_{\sigma\tau}
|(j_1 j_2)J\rangle \hspace*{5.5cm}\nonumber\\
= - \sum_L 3(2L+1)\,{{J(J+1)-L(L+1)-2}\over 4}
 \sqrt{(2j_1+1)(2j_2+1)(2j'_1+1)(2j'_2+1)} \nonumber\\
\times \left\{\begin{array}{ccc}l_1&{1\over 2}&j_1\\ l_2&{1\over 2}&j_2\\
L&1&J\end{array}\right\}
\left\{\begin{array}{ccc}l'_1&{1\over 2}&j'_1\\ l'_2&{1\over 2}&j'_2\\
L&1&J\end{array}\right\}
\langle (l'_1 l'_2)L| f_\mu^{\rm LS} (r_{12})|(l_1 l_2)L\rangle\,
\langle{\cal O}_{\sigma\tau}\rangle_{S=1} \,.\quad
\end{eqnarray}
The $\langle (l'_1 l'_2)L| f_\mu^{\rm LS} (r_{12})|(l_1 l_2)L\rangle$
matrix elements are given in Eq.~(\ref{eq:tbcrad}),
except that $\tilde f_\mu^{\rm C}(k)$
is replaced by $\tilde f_\mu^{\rm LS}(k)$,
the Fourier transform of $f_\mu^{\rm LS}(r_{12})$.

For the part including $({\mathbf r}_1\times{\mathbf p}_2)$,
we separate the spatial and the spin parts again, having
\begin{eqnarray}
\langle (j'_1 j'_2)J| f_\mu^{\rm LS} (r_{12})
\,{1\over 2}({\mathbf r}_1\times{\mathbf p}_2)
\cdot({\mathbf s}_1+{\mathbf s}_2) {\cal O}_{\sigma\tau}
|(j_1 j_2)J\rangle \hspace*{5.5cm}\nonumber\\
= - \sum_{L,L'} 3 \sqrt{6(2L+1)(2L'+1)
(2j_1+1)(2j_2+1)(2j'_1+1)(2j'_2+1)}\, W(L\,J\,1\,1;1\,L')\nonumber\\
\times
\left\{\begin{array}{ccc}l_1&{1\over 2}&j_1\\ l_2&{1\over 2}&j_2\\
L&1&J\end{array}\right\}
\left\{\begin{array}{ccc}l'_1&{1\over 2}&j'_1\\ l'_2&{1\over 2}&j'_2\\
L'&1&J\end{array}\right\}
\langle (l'_1 l'_2)L'|| f_\mu^{\rm LS} (r_{12})
\,{1\over 2}({\mathbf r}_1\times{\mathbf p}_2)^{(1)}||(l_1 l_2)L\rangle\,
\nonumber\\
\times \langle{\cal O}_{\sigma\tau}\rangle_{S=1} \,,
\end{eqnarray}
As has been shown for the central part in Section~\ref{sec:tbme},
$f_\mu^{\rm LS}(r_{12})$ contains the angular part
$[Y^{(\lambda)}(\hat{\mathbf r}_1)\cdot
Y^{(\lambda)}(\hat{\mathbf r}_2)]$.
Combining it with the $({\mathbf r}_1\times{\mathbf p}_2)$ operator,
we obtain
\begin{eqnarray}
\left[Y^{(\lambda)}(\hat{\mathbf r}_1)\cdot
Y^{(\lambda)}(\hat{\mathbf r}_2)\right]\,
{1\over 2}({\mathbf r}_1\times{\mathbf p}_2)^{(1)}
= (-)^{\lambda+1} \sqrt{{2\lambda+1}\over 2}\,
\left[Y^{(\lambda)}(\hat{\mathbf r}_1)\,
Y^{(\lambda)}(\hat{\mathbf r}_2)\right]^{(0)}\,
\left[r_1^{(1)}\nabla_2^{(1)}\right]^{(1)} \nonumber\\
= (-)^{\lambda+1} \sum_{\lambda_1,\lambda_2}
\sqrt{{(2\lambda+1)(2\lambda_2+1)}\over 2}\,
(\lambda\,0\,1\,0|\lambda_1\,0)\,
W(1\,1\,\lambda_1\,\lambda; 1\,\lambda_2) \hspace*{3.5cm}\nonumber\\
\times r_1 \left\{Y^{(\lambda_1)}(\hat{\mathbf r}_1)
\left[Y^{(\lambda)}(\hat{\mathbf r}_2)\nabla_2^{(1)}\right]^{(\lambda_2)}
\right\}^{(1)}\,.\quad
\end{eqnarray}
The Fourier transformation of Eq.~(\ref{eq:Four1})
and the integration of the angular part yields
\begin{eqnarray}
\langle (l'_1 l'_2)L'|| f_\mu^{\rm LS} (r_{12})
\,{1\over 2}({\mathbf r}_1\times{\mathbf p}_2)^{(1)}||(l_1 l_2)L\rangle
\hspace*{8cm}\nonumber\\
= \sum_{\lambda_1,\lambda_2} (-)^{\lambda_2}(2\lambda_2+1)
\sqrt{{3(2\lambda_1+1)(2L+1)(2L'+1)}\over 2}\,
(\lambda\,0\,1\,0|\lambda_1\,0)\,
W(1\,1\,\lambda_1\,\lambda; 1\,\lambda_2)\, \nonumber\\
\times \left\{\begin{array}{ccc}l_1&l_2&L\\ \lambda_1&\lambda_2&1\\
l'_1&l'_2&L'\end{array}\right\}
\cdot \int_0^\infty k^2dk\,\tilde f_\mu^{\rm LS}(k)
\sqrt{2l_1+1}\,(l_1\,0\,\lambda_1\,0|l'_1\,0)
\,{\cal I}_1^{(1)}(k) \hspace*{1cm}\nonumber\\
\times \left\{ \sqrt{(l_2+1)(2l_2+3)}\,
(l_2\!+\!1\,0\,\lambda\,0|l'_2\,0)\,
W(l_2\,1\,l'_2\,\lambda; l_2\!+\!1\,\lambda_2)\,
{\cal I}_2^{(d+)}(k) \right. \hspace*{1.5cm}\nonumber\\
\left. - \sqrt{l_2(2l_2-1)}\,
(l_2\!-\!1\,0\,\lambda\,0|l'_2\,0)\,
W(l_2\,1\,l'_2\,\lambda; l_2\!-\!1\,\lambda_2)\,
{\cal I}_2^{(d-)}(k) \right\}\,, \hspace*{1.5cm}
\end{eqnarray}
where
\begin{eqnarray}
{\cal I}_i^{(1)}(k) &=&
\int_0^\infty r^2 dr\,r j_\lambda(kr) R_{j'_i}(r) R_{j_i}(r)
\,,\nonumber\\
{\cal I}_i^{(d+)}(k) &=&
\int_0^\infty r^2 dr\, j_\lambda(kr) R_{j'_i}(r)
\left[\left({d\over{dr}}-{\ell_i\over r}\right) R_{j_i}(r)\right]
\,,\nonumber\\
{\cal I}_i^{(d-)}(k) &=&
\int_0^\infty r^2 dr\, j_\lambda(kr) R_{j'_i}(r)
\left[\left({d\over{dr}}+{{\ell_i+1}\over r}\right) R_{j_i}(r)\right]
\,.
\label{eq:r-int-LS}\end{eqnarray}
The subscript $i$ to ${\cal I}$ corresponds to the nucleon index.
The Gaussian bases of Eq.~(\ref{eq:basis}) give
\begin{eqnarray}
\left({d\over{dr}}-{\ell\over r}\right) R_{\alpha\ell j}(r)
&=& \left({{2p_\alpha}\over{r}} - {{2r}\over{\nu_\alpha^2}} \right)
R_{\alpha\ell j}(r) \,,\nonumber\\
\left({d\over{dr}}+{{\ell+1}\over r}\right) R_{\alpha\ell j}(r)
&=& \left({{2\ell+2p_\alpha+1}\over{r}} - {{2r}\over{\nu_\alpha^2}} \right)
R_{\alpha\ell j}(r) \,.
\end{eqnarray}
Hence the $r$ integration in Eq.~(\ref{eq:r-int-LS})
is implemented in a similar manner to Eq.~(\ref{eq:r-int-cent}).
Because of the parity selection rule,
the result is a product of a polynomial of $k$ and a Gaussian,
and finally the $k$ integration has the form of Eq.~(\ref{eq:k-integ}).
Thus the $k$ integration is carried out analytically
for $f_\mu^{\rm LS}(r_{12}) = e^{-(\mu r_{12})^2}$,
while it is expressed by using the error function
for $f_\mu^{\rm LS}(r_{12}) = e^{-\mu r_{12}}/\mu r_{12}$.
The matrix elements of the part
coming from $({\mathbf r}_2\times{\mathbf p}_1)$ in Eq.~(\ref{eq:L-sep})
are obtained by interchanging the nucleon indices 1 and 2,
with an appropriate phase factor.

In the Skyrme and the Gogny interactions,
the LS part is taken to be
\begin{equation} v_{12}^{\rm LS} =
2iW_0 \left[{\mathbf p}_{12} \times \delta({\mathbf r}_{12})
{\mathbf p}_{12}\right] \cdot ({\mathbf s}_1+{\mathbf s}_2) \,.
\label{eq:Sky-LS1}\end{equation}
This type of the LS interaction is also treated
in an analogous manner, by using the Fourier transformation.
It is noted that this $v_{12}^{\rm LS}$ force
operates only on the $S=1$, $T=1$ ({\it i.e.} spatially odd) channel,
because of the $({\mathbf s}_1+{\mathbf s}_2)$ operator
and of $\delta({\mathbf r}_{12})$.
Recall that
\begin{equation}
\delta({\mathbf r}_{12}) = \lim_{\mu\rightarrow\infty}
\left({\mu\over\sqrt\pi}\right)^3\,e^{-(\mu r_{12})^2}\,.
\label{eq:delta-lim}\end{equation}
Substituting Eq.~(\ref{eq:delta-lim}) into Eq.~(\ref{eq:Sky-LS1}),
we obtain
\begin{eqnarray} v_{12}^{\rm LS} &=&
2iW_0 \lim_{\mu\rightarrow\infty} \left({\mu\over\sqrt\pi}\right)^3
\left[{\mathbf p}_{12} \times e^{-(\mu r_{12})^2}
{\mathbf p}_{12}\right] \cdot ({\mathbf s}_1+{\mathbf s}_2) \nonumber\\
&=& -4W_0 \left[\lim_{\mu\rightarrow\infty}
{{\mu^5}\over{\pi^{3\over2}}} e^{-(\mu r_{12})^2}\right]
{\mathbf L}_{12}\cdot ({\mathbf s}_1+{\mathbf s}_2) \,.
\label{eq:Sky-LS3}\end{eqnarray}
We shall take the $\mu\rightarrow\infty$ limit after a certain algebra.
By comparing Eq.~(\ref{eq:Sky-LS3}) with Eq.~(\ref{eq:effint}),
we can identify
\begin{equation} f_{\delta''}^{\rm LS}(r_{12})
= -4 \lim_{\mu\rightarrow\infty}
{{\mu^5}\over{\pi^{3\over2}}} e^{-(\mu r_{12})^2}\,,
\label{eq:f_del1}\end{equation}
whose Fourier transform is
\begin{equation} \tilde f_{\delta''}^{\rm LS}(k)
= -4 \lim_{\mu\rightarrow\infty} \mu^2\,e^{-(k/2\mu)^2}\,,
\end{equation}
and $t_{\delta''}^{\rm LSO}=W_0$.
We here use the suffix $\delta''$ to stand for the spatial part
of the LS force (\ref{eq:Sky-LS1}),
for the reason clarified in the discussion below.
By expanding the exponential factor
and taking the $\mu\rightarrow\infty$ limit
except the non-vanishing terms, we have
\begin{equation} \tilde f_{\delta''}^{\rm LS}(k)
= -4 \left(\lim_{\mu\rightarrow\infty} \mu^2\right) + k^2\,.
\end{equation}
Although the first term in the right-hand side looks divergent,
it is obvious that its inverse Fourier transform
is proportional to $\delta({\mathbf r}_{12})$,
which turns out to vanish
because $\delta({\mathbf r}_{12})\,{\mathbf L}_{12}=0$.
We now pose safely, for the LS force of Eq.~(\ref{eq:Sky-LS1}),
\begin{equation} \tilde f_{\delta''}^{\rm LS}(k) = k^2 \,.
\label{eq:LS-Four}\end{equation}
This is the same form as in the momentum-dependent term
discussed in Appendix~\ref{app:mom-dep}.
Equation~(\ref{eq:LS-Four}) corresponds to the modification of
Eq.~(\ref{eq:f_del1}) as
\begin{equation} f_{\delta''}^{\rm LS}(r_{12}) =
-4 \lim_{\mu\rightarrow\infty} \mu^2 \left[
\left({\mu\over\sqrt\pi}\right)^3\,e^{-(\mu r_{12})^2}
- \delta({\mathbf r}_{12})\right]\,.
\end{equation}
Thus the LS force in the Skyrme and Gogny interactions
is treated in a unified way with the finite-range LS forces.

\section{Matrix elements of tensor interaction}
\label{app:tensor}

We turn to the non-anti-symmetrized matrix elements
of the tensor force,
\begin{equation}
\langle (j'_1 j'_2)J| v_{12}^{\rm TN} |(j_1 j_2)J\rangle
= \sum_\mu \langle (j'_1 j'_2)J| (t_\mu^{\rm TNE} P_{\rm TE}
 + t_\mu^{\rm TNO} P_{\rm TO}) f_\mu^{\rm TN} (r_{12})
\,r_{12}^2 S_{12} |(j_1 j_2)J\rangle\,.
\label{eq:tbTN}\end{equation}
The tensor force operates only on the spin-triplet two-particle states.
As in Appendix~\ref{app:LS}, we shall denote $P_{\rm TE}$ or $P_{\rm TO}$
by ${\cal O}_{\sigma\tau}$.

By separating the spatial degrees-of-freedom from the spin ones,
the tensor operator of Eq.~(\ref{eq:tensor}) is rewritten as
\begin{equation}
r_{12}^2 S_{12} = 8\left\{
\sqrt{{6\pi}\over 5}\left[r_1^2 Y^{(2)}(\hat{\mathbf r}_1)
+ r_2^2 Y^{(2)}(\hat{\mathbf r}_2)\right]
- 4\pi r_1 r_2 \left[Y^{(1)}(\hat{\mathbf r}_1)
Y^{(1)}(\hat{\mathbf r}_2)\right]^{(2)} \right\}
\cdot \left[s_1^{(1)} s_2^{(1)}\right]^{(2)}\,.
\label{eq:TN-sep}\end{equation}
The matrix element of the $r_1^2 Y^{(2)}(\hat{\mathbf r}_1)$ part
is given by, after calculating the spin part,
\begin{eqnarray}
\langle (j'_1 j'_2)J| f_\mu^{\rm TN} (r_{12})\cdot
8\sqrt{{6\pi}\over 5}\,r_1^2 Y^{(2)}(\hat{\mathbf r}_1)
\cdot \left[s_1^{(1)} s_2^{(1)}\right]^{(2)}
{\cal O}_{\sigma\tau} |(j_1 j_2)J\rangle \hspace*{3.5cm}\nonumber\\
= \sum_{L,L'} 12 \sqrt{5(2L+1)(2L'+1)
(2j_1+1)(2j_2+1)(2j'_1+1)(2j'_2+1)}\,W(L\,J\,2\,1;1\,L')\nonumber\\
\times
\left\{\begin{array}{ccc}l_1&{1\over 2}&j_1\\ l_2&{1\over 2}&j_2\\
L&1&J\end{array}\right\}
\left\{\begin{array}{ccc}l'_1&{1\over 2}&j'_1\\ l'_2&{1\over 2}&j'_2\\
L'&1&J\end{array}\right\}
\langle (l'_1 l'_2)L'|| f_\mu^{\rm TN}(r_{12})
\sqrt{{6\pi}\over 5}r_1^2 Y^{(2)}(\hat{\mathbf r}_1)
||(l_1 l_2)L\rangle \nonumber\\
\times \langle{\cal O}_{\sigma\tau}\rangle_{S=1} \,.\quad
\end{eqnarray}
By using the Fourier transform of $f_\mu^{\rm TN}(r_{12})$
and integrating out the angular part,
we obtain for the spatial matrix element,
\begin{eqnarray}
\langle (l'_1 l'_2)L'|| f_\mu^{\rm TN}(r_{12})
\sqrt{{6\pi}\over 5}r_1^2 Y^{(2)}(\hat{\mathbf r}_1)
||(l_1 l_2)L\rangle \hspace*{7cm}\nonumber\\
= (-)^\lambda \sum_{\lambda_1}
\sqrt{{3(2\lambda_1+1)(2L+1)(2L'+1)}\over 2}
\,(\lambda\,0\,2\,0|\lambda_1\,0)
\left\{\begin{array}{ccc}l_1&l_2&L\\ \lambda_1&\lambda&2\\
l'_1&l'_2&L'\end{array}\right\} \hspace*{2cm}\nonumber\\
\times \int_0^\infty k^2dk\,\tilde f_\mu^{\rm TN}(k) \sqrt{2l_1+1}
\,(l_1\,0\,\lambda_1\,0|l'_1\,0)\,{\cal I}^{(2)}_1(k)
\cdot \sqrt{2l_2+1}
\,(l_2\,0\,\lambda\,0|l'_2\,0)\,{\cal I}^{(0)}_2(k)\,,
\end{eqnarray}
where
\begin{equation}
{\cal I}_i^{(2)}(k) = \int_0^\infty r^2dr\,
r^2 j_\lambda(kr) R_{j'_i}(r) R_{j_i}(r)\,,
\end{equation}
and ${\cal I}_i^{(0)}$ is defined in Eq.~(\ref{eq:r-int-cent0}).
The matrix element of the $r_2 Y^{(2)}(\hat{\mathbf r}_2)$ part
in the expansion of Eq.~(\ref{eq:TN-sep})
is obtained by an appropriate replacement
between the nucleon indices.

After the algebra regarding the spin degrees-of-freedom,
we have for the matrix elements of the $r_1 r_2
[Y^{(1)}(\hat{\mathbf r}_1) Y^{(1)}(\hat{\mathbf r}_2)]^{(2)}$ part,
\begin{eqnarray}
\langle (j'_1 j'_2)J| f_\mu^{\rm TN} (r_{12})\cdot
8\cdot 4\pi\,r_1 r_2 \left[Y^{(1)}(\hat{\mathbf r}_1)
Y^{(1)}(\hat{\mathbf r}_2)\right]^{(2)}
\cdot \left[s_1^{(1)} s_2^{(1)}\right]^{(2)}
{\cal O}_{\sigma\tau} |(j_1 j_2)J\rangle
\hspace*{1.5cm}\nonumber\\
= \sum_{L,L'} 12 \sqrt{5(2L+1)(2L'+1)
(2j_1+1)(2j_2+1)(2j'_1+1)(2j'_2+1)}\,W(L\,J\,2\,1;1\,L')
\hspace*{1cm}\nonumber\\
\times
\left\{\begin{array}{ccc}l_1&{1\over 2}&j_1\\ l_2&{1\over 2}&j_2\\
L&1&J\end{array}\right\}
\left\{\begin{array}{ccc}l'_1&{1\over 2}&j'_1\\ l'_2&{1\over 2}&j'_2\\
L'&1&J\end{array}\right\}
\langle (l'_1 l'_2)L'|| f_\mu^{\rm TN}(r_{12})\,4\pi
r_1 r_2 \left[Y^{(1)}(\hat{\mathbf r}_1)
Y^{(1)}(\hat{\mathbf r}_2)\right]^{(2)}
||(l_1 l_2)L\rangle \nonumber\\
\times \langle{\cal O}_{\sigma\tau}\rangle_{S=1} \,.\quad
\end{eqnarray}
For the spatial matrix elements, we obtain
\begin{eqnarray}
\langle (l'_1 l'_2)L'|| f_\mu^{\rm TN}(r_{12})
\,4\pi r_1 r_2 \left[Y^{(1)}(\hat{\mathbf r}_1)
Y^{(1)}(\hat{\mathbf r}_2)\right]^{(2)}
||(l_1 l_2)L\rangle \hspace*{5cm}\nonumber\\
= (-)^\lambda \sum_{\lambda_1,\lambda_2}
3\sqrt{5(2\lambda_1+1)(2\lambda_2+1)(2L+1)(2L'+1)}
\,(\lambda\,0\,1\,0|\lambda_1\,0)
\,(\lambda\,0\,1\,0|\lambda_2\,0) \hspace*{0.5cm}\nonumber\\
\times W(2\,1\,\lambda_1\,\lambda;1\,\lambda_2)
\left\{\begin{array}{ccc}l_1&l_2&L\\ \lambda_1&\lambda_2&2\\
l'_1&l'_2&L'\end{array}\right\}
\int_0^\infty k^2dk \sqrt{2l_1+1}
\,(l_1\,0\,\lambda_1\,0|l'_1\,0)\,{\cal I}^{(1)}_1(k) \nonumber\\
\times \sqrt{2l_2+1}
\,(l_2\,0\,\lambda_2\,0|l'_2\,0)\,{\cal I}^{(1)}_2(k)\,,
\end{eqnarray}
where ${\cal I}_i^{(1)}$ is given in Eq.~(\ref{eq:r-int-LS}).

In Ref.~\cite{ref:S3}, a zero-range tensor interaction
was proposed to cancel out a certain term of the LS current.
By separating it into the spatially even and odd channels,
we have
\begin{eqnarray} v_{12}^{\rm TN} =
4t_{\delta''''}^{\rm TNE}
\left\{\left[3({\mathbf s}_1\cdot{\mathbf p}_{12})
({\mathbf s}_2\cdot{\mathbf p}_{12})
- ({\mathbf s}_1\cdot{\mathbf s}_2){\mathbf p}_{12}^2\right]
\delta({\mathbf r}_{12})\right. \hspace*{2.5cm}\nonumber\\
\left. + \delta({\mathbf r}_{12})
\left[3({\mathbf s}_1\cdot{\mathbf p}_{12})
({\mathbf s}_2\cdot{\mathbf p}_{12})
- ({\mathbf s}_1\cdot{\mathbf s}_2){\mathbf p}_{12}^2\right]\right\}
P_{\rm TE} \nonumber\\
-4\,t_{\delta''''}^{\rm TNO}\left\{
3({\mathbf s}_1\cdot{\mathbf p}_{12})\delta({\mathbf r}_{12})
({\mathbf s}_2\cdot{\mathbf p}_{12})
+ 3({\mathbf s}_2\cdot{\mathbf p}_{12})\delta({\mathbf r}_{12})
({\mathbf s}_1\cdot{\mathbf p}_{12})\right. \nonumber\\
\left. - 2({\mathbf s}_1\cdot{\mathbf s}_2)
\left[{\mathbf p}_{12}\cdot\delta({\mathbf r}_{12})
{\mathbf p}_{12}\right]\right\} P_{\rm TO}\,.
\end{eqnarray}
In Ref.~\cite{ref:S3} $t_{\delta''''}^{\rm TNO}
={3\over 8}t_{\delta''''}^{\rm TNE}$.
By using the expression of Eq.~(\ref{eq:delta-lim}),
we obtain
\begin{equation}
f_{\delta''''}^{\rm TN}(r_{12}) = \lim_{\mu\rightarrow\infty}
{{4\mu^7}\over{\pi^{3\over 2}}}\,e^{-(\mu r_{12})^2}\,,
\end{equation}
whose Fourier transform yields
\begin{equation}
\tilde f_{\delta''''}^{\rm TN}(k) = {k^4\over 8}\,,
\end{equation}
after eliminating the vanishing terms.
Thus this type of the tensor force is also treated
in a unified way with the finite-range tensor force.

\clearpage
\begin{table}
\begin{center}
\caption{Hartree-Fock energies (MeV) in the D1S interaction
for $^{16}$O, $^{24}$O and $^{28}$O,
using the HO, KG and hybrid basis-sets.
\label{tab:eng_D1S}}
\begin{tabular}{crrr}
\hline\hline
nuclide & HO~~ & KG~~ & hybrid \\ \hline
$^{16}$O & $-136.849$ & $-136.720$ & $-136.952$ \\
$^{24}$O & $-179.477$ & $-179.445$ & $-179.815$ \\
$^{28}$O & $-181.397$ & $-181.501$ & $-181.770$ \\
\hline\hline
\end{tabular}
\end{center}
\end{table}

\clearpage

\end{document}